\documentclass[fleqn,usenatbib]{mnras}

\usepackage{newtxtext,newtxmath}

\usepackage[T1]{fontenc}

\DeclareRobustCommand{\VAN}[3]{#2}
\let\VANthebibliography\thebibliography
\def\thebibliography{\DeclareRobustCommand{\VAN}[3]{##3}\VANthebibliography}

\usepackage{graphicx}	
\usepackage{amsmath}	

\title[Simulated tidal and secular bars]{Origins and lifetimes of secular and tidal bars in simulated disc galaxies}

\author[M. Frosst]{
Matthew Frosst$^{1,2}$\thanks{E-mail: matt.frosst@icrar.org},
Danail Obreschkow$^{1,2}$,
Aaron Ludlow$^{1,2}$,
Amelia Fraser-McKelvie$^{2,3}$
\\
$^{1}$International Centre for Radio Astronomy Research (ICRAR), University of Western Australia, Crawley, WA 6009, Australia\\
$^{2}$ARC Centre of Excellence for All Sky Astrophysics in 3 Dimensions (ASTRO 3D), Sydney, Australia\\
$^{3}$European Southern Observatory, Karl-Schwarzschild-Straße 2, Garching, 85748, Germany
}

\date{Accepted XXX. Received YYY; in original form ZZZ}

\pubyear{2023}

\begin{document}
\label{firstpage}
\pagerange{\pageref{firstpage}--\pageref{lastpage}}
\maketitle

\begin{abstract}
We investigate the formation of stellar bars in $307$ Milky Way-mass disc galaxies in the \textsc{TNG50} cosmological simulation. Most bars form rapidly in dynamically cold discs shortly after the central stellar mass exceeds that of dark matter. In these cases, bar formation is consistent with secular instabilities driven by the disc’s self-gravity, which organises stellar orbits into a coherent bar structure. However, around $25$ per cent of barred galaxies are dark matter dominated at the time of bar formation, $t_{\rm bar}$, and remain so thereafter. We trace the origin of these bars to tidal perturbations induced by passing satellites or mergers using a new metric, $\mathcal{S}_{\rm bar}$, quantifying the tidal field acting on the galaxy. At the time of bar formation, we find a negative correlation between $\mathcal{S}_{\rm bar}$ and the central stellar-to-dark matter mass fraction, indicating that more dark matter-dominated discs require stronger tides to trigger bar formation. These tidally induced bars are more likely to be transient than those that form secularly, although bar properties are otherwise similar. However, the host galaxies differ: secular bars arise in relatively compact discs, while tidal bars appear in extended discs whose properties resemble those of unbarred galaxies. Tidal perturbations can therefore induce bars in galaxies otherwise stable to secular formation, highlighting the dual role of the internal galactic structure and the external environment in bar formation.
\end{abstract}



\section{Introduction}
The recent explosion in observations of high-redshift discs has revealed an early Universe populated by settled, disc-like galaxies \citep{Ferreira2023, Conselice2024}. Stellar bars, too, are now regularly observed out to $z \approx 1-3$ \citep[e.g.][]{Guo2023,LeConte2024,Guo2024,EspejoSalcedo2025}, indicating that bar formation is not restricted to low$-z$ discs. They are believed to form via two processes: the secular growth of a global disc instability \citep{Hohl1971, ELN1982, Fujii2018, Bland-Hawthorn2024} and external tidal interactions that are common in cosmological environments \citep{Noguchi1987,Berentzen2004,Lokas2014}. Despite their ubiquity, it remains uncertain whether bars that form secularly can be distinguished from those that form as a result of tidal interactions, particularly long after their formation \citep{Cavanagh2022, Iles2024, Zheng2025}. 
 
Numerical simulations show that bars form secularly in dynamically cold, self-gravitating discs with high stellar-to-dark matter (DM) mass ratios \citep[e.g.,][]{Fujii2018, BlandHawthorn2023, Sellwood2014, Frosst2024}, while DM-dominated galaxies may not form bars. Other disc properties, like thickness \citep[e.g.][]{Klypin2009, Aumer2017, Ghosh2022}, central velocity dispersion \citep[e.g.][]{Sellwood1984, Athanassoula2003}, and gas fraction \citep{Berentzen2004, Bournaud2005}, can influence the onset and speed of secular bar growth. Once formed, secular bars tend to persist for many Gyr \citep[e.g.][]{Athanassoula2013}. 

However, galaxies are subject to frequent interactions \citep{Hammer2009}. Simulations show that stellar bars can form rapidly in response to the associated tidal perturbations, particularly during mergers \citep[e.g.,][]{Noguchi1987, Berentzen2004} or flybys \citep[e.g.,][]{Lang2014, Lokas2014, Lokas2018}, during which strong tidal forces act on the disc \citep[see also][]{Mayer2004, RomanoDiaz2008, Martinez2017, Zana2018a, Peschken2019, RosasGuevara2024}. These perturbations can lead to bar formation even in galaxies that are stable to secular bar growth \citep{Noguchi1987, RomanoDiaz2008, Lang2014, Zheng2025}. Furthermore, tidal bars may be short-lived and prone to destruction \citep[][]{Peschken2019}. Cosmological simulations reveal a wide range of bar lifetimes, with many galaxies experiencing multiple episodes of bar formation, termination, and regeneration \citep{Berentzen2004,RomanoDiaz2008, Ansar2023}.

Whether tidally-induced and secular bars have distinct characteristics is an open question. Early simulations by \citet{Noguchi1987} suggest that tidal bars resemble those that formed secularly, exhibiting similar lengths, strengths, and pattern speeds \citep[see also][]{Moetazedian2017, Zana2018a}. However, later studies reported that tidally-induced bars are stronger \citep[e.g.][]{Miwa1998, Lokas2014, Lokas2018, Peschken2019} and rotate more slowly than their secular counterparts \citep{Pettitt2018, Martinez2017}. \citet{Zheng2025} reconciled this difference by using idealised simulations to show that slowly rotating tidal bars tend to form in galaxies that are otherwise stable against secular bar formation. In contrast, they show that tidal bars can also rotate as rapidly as secular bars, but only in galaxies that could \textit{also} have formed the bar secularly \citep[see also][]{Martinez2017}. This again highlights the importance of disc properties in modulating the impact of tidal bar formation. Given what may be stark differences, is there a reliable way to distinguish between secular and tidal bars after they have formed?

Most previous studies comparing tidal and secular bars were based on idealised \citep[e.g.][]{Lang2014, Lokas2018, Iles2024, Zheng2025} or zoom-in simulations \citep[e.g.][]{Zana2018a, Zana2018b, Ansar2023}. While these approaches allow for high spatial and mass resolution (and controlled environments in the case of idealised runs), they do not capture the diversity of galaxy-galaxy interactions that may affect the properties of tidal bars \citep{Curir2006}. 

Cosmological simulations of large volumes now reach sufficient resolution to reliably model the formation and evolution of stellar bars in statistically meaningful numbers of Milky Way-mass (MW) galaxies, the mass regime at which bars are most commonly found \citep{Melvin2014, Gavazzi2015}. For instance, \citet{RosasGuevara2024} used the \textsc{TNG50} simulation to study the transformation of disc galaxies between $z=0$ and $1$, and found that roughly one-third of all bar formation events were tidally driven. \citet[][]{Lopez2024} studied a smaller sample of disc galaxies in \textsc{TNG50} and found that while barred galaxies inhabit denser environments, it was challenging to link bar formation to previous mergers or flybys. In contrast, \citet{Peschken2019} analysed stellar bars in the \textsc{Illustris} simulation, and found that nearly all were tidally induced; this is related to their lower resolution. The different conclusions of these studies expose the uncertainty surrounding environmental triggers of bar formation. 

To address this uncertainty, we will analyse a sample of 307 $z=0$ disc galaxies with stellar masses comparable to that of the Milky Way, drawn from the \textsc{TNG50} simulation. This mass range is well-studied in the context of bar formation \citep[e.g.][]{RosasGuevara2022, Khoperskov2024, Pillepich2024, Semczuk2024, Frosst2025}, and sufficiently well resolved for robust bar evolution \citep[e.g.][]{Debattista2008, Frosst2024}. By following these galaxies from $z=4$ to $0$, we will investigate how disc properties and tidal interactions influence stellar bar formation, and whether tidally induced and secular bars can be distinguished. 

This paper is organised as follows: in Section~\ref{sec:methods}, we describe the \textsc{TNG50} simulation and the selection criteria for our galaxy sample. Section~\ref{sec:baranalysis} details our methods for characterising bar properties and quantifying the frequency of bar formation episodes. In Section~\ref{sec:tides}, we introduce a metric to quantify the tidal fields acting on galaxies and use it to characterise galactic tides near the time of bar formation. In Section~\ref{sec:results}, we present our results on how galaxy and bar properties are related to the underlying formation mechanisms. We summarise our findings in Section~\ref{sec:conclusions}.

\section{Methodology}\label{sec:methods}
Throughout the paper, we define the stellar mass of a galaxy, ${\rm M}_{\star}$, as the total mass of stellar particles bounded by a $r=30\,{\rm kpc}$ sphere that is centred on the particle with the minimum potential energy, which we define as the galaxy centre. We adopt a cylindrical coordinate system coincident with the galaxy centre and align the $z$-axis with the net angular momentum vector of all stellar particles and star-forming gas cells within two stellar half mass radii, i.e. $2\times r_{\star,1/2}$. 
In this coordinate system, $r = (R^2 + z^2)^{1/2}$ denotes the three-dimensional radius, where $R=(x^2+y^2)^{1/2}$ is the distance from the $z$-axis. 

\subsection{The \textsc{TNG50} simulation}\label{sec:code} 
TNG50-1 \citep[TNG50 hereafter;][]{Pillepich2018a, Nelson2019a} is the highest resolution simulation of the TNG project \citep{Marinacci2018, Naiman2018, Nelson2018, Springel2018}. It was run using the \textsc{AREPO} moving-mesh code \citep{Springel2010} with cosmological parameters taken from the \citet{Planck2016} results. 

\textsc{TNG50} follows the evolution of DM, gas cells, and stellar and black hole particles in a $51.7$ cubic cMpc volume from $z=127$ to $z=0$ (the prefix ``c'' indicates co-moving Mpc). At the initial redshift, the simulation contains $2160^3$ fluid elements that have a target mass of $m_{\rm gas}=8.5\times10^4{\rm M}_\odot$ and an equal number of DM particles of mass $m_{\rm DM}=4.5\times10^5{\, \rm M_{\odot}}$. \textsc{TNG50} employs subgrid models for unresolved physics including heating and cooling, star formation and stellar evolution, chemical enrichment, supermassive black hole seeding and growth, and feedback from stars and active galactic nuclei \citep[see][for details]{Pillepich2018b, Springel2018, Nelson2018, Naiman2018, Marinacci2018}. The subgrid model parameters were calibrated as described in \citet{Weinberger2017}.

The gravitational softening for gas cells is dynamically adapted to the effective cell size, but reaches a minimum physical value of $\epsilon_{\rm gas} = 72$ pc. The softening length for collisionless (DM and stellar) particles is $\epsilon_{c} = 575$ cpc until $z=1$, and is fixed at $\epsilon_{c} = 288$ pc at lower redshifts.

Galaxies and DM haloes were identified using the \textsc{SUBFIND} algorithm \citep{Springel2001, Dolag2009} and linked between consecutive snapshots using the \textsc{Sublink} merger tree code \citep{RodriguezGomez2015}; this also returns the virial properties of the halo, for instance, $M_{\rm 200c}$, the mass within the radius $r_{\rm 200c}$ that encloses a mean density equal to $200\times$ the critical density of the universe, $\rho_{\rm crit}$.

\subsection{The disc galaxy sample}\label{sec:sample}

We focus our analysis on a sample of well-resolved galaxies that host prominent discs at $z=0$. We characterise the morphologies of galaxies using $\kappa_\star$, i.e. the fraction of stellar kinetic energy in ordered rotation, defined by \citet{Sales2010} as
\begin{equation}\label{eq:kappa}
    \kappa_{\star} = \frac{\sum_{k} m_{k}(j_{z,k} / R_{k})^{2}}{\sum_{k}m_{k}v_{k}^{2}},
\end{equation}
where $j_{z,k}$, $v_{k}$, and $m_{k}$ are the $z$-component of the specific angular momentum, velocity magnitude, and mass of the $k^{\rm th}$ particle, respectively, and the sum extends over all stellar particles with $r_k\leq 3r_{\star,1/2}$. Galaxies with prominent stellar discs have higher values of $\kappa_{\star}$ than dispersion-dominated spheroids \citep[e.g.][]{Correa2017, RodriguezGomez2017}. 

The sample of disc galaxies used in our analysis includes all 307 disc galaxies in \textsc{TNG50} that satisfy the following criteria at $z = 0$: 
\begin{enumerate} 
\item A stellar mass, $M{\star}$, in the range $10^{10.5}{\rm M{\odot}} < M{\star} \leq 10^{11.2}{\rm M{\odot}}$; 
\item $\kappa_{\star} \geq 0.3$; 
\item No other galaxy with $M_{\star} \geq 10^{10}{\rm M_{\odot}}$ lies within 100 kpc of their centers. 
\end{enumerate} 
This sample includes the 198 MW/M31 analogs curated by \citet{Pillepich2024}, and extends it to a broader range of environments. 

\section{Bar formation and evolution}\label{sec:baranalysis}
\subsection{Characterising the formation of stellar bars}
We determine the presence of stellar bars using a Fourier decomposition of the two-dimensional, face-on stellar surface mass density of each disc \citep[][]{Athanassoula2002a, Guo2019, Dehnen2023, Frosst2024}. Specifically, we focus on the $m=2$ cylindrical mode, defined as
\begin{equation}
    \mathcal{A_{\rm 2}} = \frac{\sum_k m_{k} e^{2i\theta_{k}}}{\sum_k m_{k}},
\end{equation}
where $m_{k}$ and $\theta_{k}$ are the mass and azimuthal angle of the $k^{\rm th}$ star particle. We use this to construct radial profiles of the bar strength,
\begin{equation} \label{eq:A2}
    A_{\rm 2}(R) = |\mathcal{A}_{\rm 2}(R)|,
\end{equation}
and its position angle,
\begin{equation} \label{eq:phi2}
    \phi_2(R)=\frac{1}{2}\arg(\mathcal{A}_2(R)).
\end{equation}

The $A_{2}(R)$ and $\phi_{2}(R)$ profiles are calculated in cylindrical shells that contain $N_{\rm part}=10^4$ stellar particles. From these two profiles, we define four bar properties:
\begin{enumerate}
    \item The characteristic bar strength, $A_{2}^{\rm max}$, defined at the radius where $A_{2}(R)$ reaches a maximum. 
    \item The bar position angle, $\phi_{2}^{\rm max}$, defined as the value of $\phi_{2(R)}$ at the same radius at which $A_{2}(R)$ reaches a maximum. 
    \item The bar length, $R_{\rm bar}$, defined as the radius of the largest shell where $A_{2}(R_{\rm bar}) \geq A_{2}^{\rm max}/2$ and $|\phi_{2}(R_{\rm bar})| \leq \phi_{2}^{\rm max} + 10^{\circ}$.
    \item The instantaneous bar pattern speed, $\Omega_{\rm bar}$, is calculated from the time derivative of $\phi_{2}^{\rm max}$ while accounting for particle flux between bins \citep[following the method outlined by][]{Dehnen2023}. 
\end{enumerate}
Further details on our measurements of bar properties can be found in Appendix~\ref{apx:bar_examples}. 

A galaxy is considered barred when $A_{2}^{\rm max} \geq 0.2$ and $R_{\rm bar} \geq 1.4 \, \epsilon_{\rm c}$ are simultaneously satisfied and persist for at least three consecutive snapshots \citep[$\sim 450\,{\rm Myr}$ for \textsc{TNG50}; see][for a similar approach]{RosasGuevara2022}. With this, we measure the bar formation time, $t_{\rm bar}$, as the start of a new bar episode, and defined as the first snapshot at which a galaxy is considered barred. We consider an existing bar episode to end if $A_{2}^{\rm max} < 0.2$ for at least three consecutive snapshots; the bar termination time corresponds to the last snapshot before this criterion is satisfied. For the termination of bars, we do not consider $\phi_2$ or $R_{\rm bar}$, as these quantities are sensitive to fluctuations induced by flybys with passing satellite galaxies, even during interactions where stellar bars are visually present. Throughout a galaxy's evolution there may be several bar episodes, and thus several formation and termination times. The bar lifetime is simply the time interval between $t_{\rm bar}$ and its termination time (or $z=0$, in which case it is a lower limit). 

\begin{figure}
	\includegraphics[width=\columnwidth]{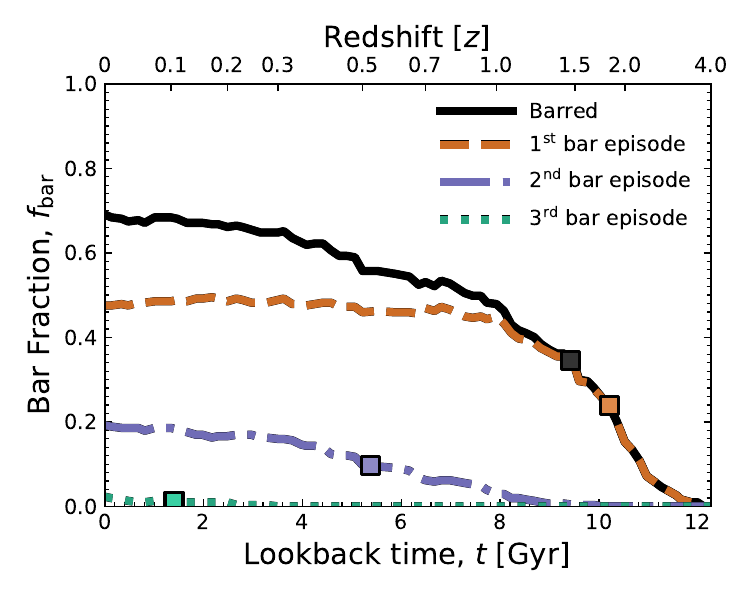}
    \caption{Evolution of the instantaneous bar fraction, $f_{\rm bar}$, for the full galaxy sample (black). The brown, purple, and green lines show the fraction of galaxies currently in their first, second, or third bar episode, respectively. Coloured squares mark the time at which these lines reach half their $z=0$ value. } 
    \label{fig:bar_fraction}
\end{figure}

\begin{figure*}
	\includegraphics[width=\textwidth]{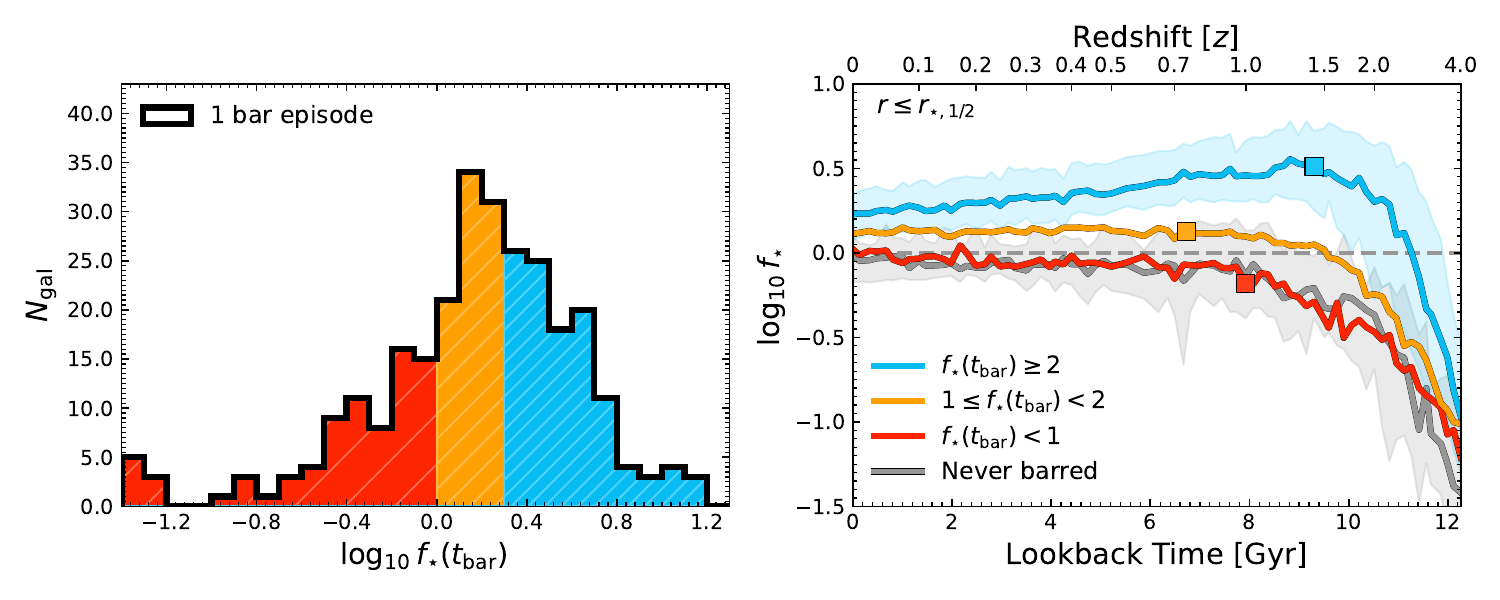}
    \caption{The left panel shows the distribution of the central stellar-to-DM mass ratio measured at the time of bar formation, $f_{\star}(t_{\rm bar})$, for all galaxies in our sample that form only one bar episode. A vertical grey dashed line indicates where $f_{\star}(t_{\rm bar}) =1$ and $=2$, from left to right, respectively. Background colours indicate the $f_{\star}(t_{\rm bar})$ bins to be used throughout this work. In the right panel, we show the evolution of $f_{\star}$ from $z=4$ to $z=0$ for each $f_{\star}(t_{\rm bar})$ bin. Coloured squares indicate the median $t_{\rm bar}$ of these bins. The grey line indicates the evolution of galaxies that never form bars. Shaded regions show the inter-quartile range (IQR) for select $f_{\star}(t_{\rm bar})$ bins, but are representative of all others.} 
    \label{fig:fdm_evolution}
\end{figure*}

\subsection{Episodes of bar formation in galactic discs} \label{sec:barfrac}
The solid black line in the top panel of Fig.~\ref{fig:bar_fraction} shows the instantaneous fraction of barred galaxies, $f_{\rm bar}$, as a function of time, $t$. The brown, purple, and green lines show the fractions of galaxies on their first, second, or third bar episode, respectively. These coloured lines add up to the total bar fraction (black line) at any time. Coloured squares along the lines mark the time at which each population reaches half its $z=0$ value. 

The instantaneous $f_{\rm bar}$ in Fig.~\ref{fig:bar_fraction} increases from about 2 per cent at $z=4$ to 69 per cent at $z=0$ (corresponding to 212 barred discs), consistent with other studies based on \textsc{TNG50} \citep{Gargiulo2022, RosasGuevara2022, Frosst2025} and with analyses of other simulations that used similar galaxy selection criteria \citep{Fragkoudi2025}. Of the galaxies in our sample that are barred at $z=0$, 69 per cent have experienced only one bar episode, with that bar persisting to the end of the simulation. Another 28 per cent are in their second bar episode, half of which have formed by $z \approx 0.5$. Only $3$ per cent have had three bar episodes, half of which have formed by $z \approx 0.15$. 

Regardless of whether the bars survive to $z=0$, when considering the full history of bars formed within our sample of discs, we find that $66$ per cent of galaxies undergo only one bar episode (203 galaxies), 24 per cent experience multiple episodes (75 galaxies), and the remaining $10$ per cent (29 galaxies) never form a bar. Among the galaxies with multiple episodes, 89 per cent form bars twice, and no galaxy experiences more than three. This upper bound may be due to the time resolution of our analysis and the relatively conservative criteria used to identify bar episodes. 

Together, these results demonstrate that, while bar formation is common in massive discs, bar histories are diverse, and that a significant population of galaxies experience only a single, sustained bar throughout their evolution. While difficult to verify observationally, this appears to be in agreement with the large number of old bars derived from age dating nuclear discs \citep{deSaFreitas2025}. For simplicity, we will focus hereafter on galaxies with only one bar episode. More complicated circumstances are left for future work. 

\subsection{The central stellar dominance of bar forming galaxies}
\citet{Fujii2018} used idealised simulations to show that secular bar formation occurs more rapidly in galaxies with higher disc-to-total mass fractions \citep[see also][]{BlandHawthorn2023}. Motivated by this, in Fig.~\ref{fig:fdm_evolution} we plot the time evolution of the central stellar-to-DM mass ratio, $f_{\star} = M_{\star}/M_{\rm DM}$, measured within $r\leq r_{\star,1/2}$, for galaxies in our sample. In principle, galaxies with high $f_{\star}$ should be able to rapidly form secular bars, but galaxies with low $f_{\star}$ should be comparatively stable against secular bar formation. 

In the left panel of Fig.~\ref{fig:fdm_evolution} we show as a thick black histogram the distribution of $f_{\star}$, at the time of bar formation, $t_{\rm bar}$, for all galaxies that form only one bar. The differently coloured regions under this line indicate whether these galaxies form their bars while highly centrally stellar-dominated ($f_{\star}(t_{\rm bar}) \geq 2$, blue, tightly hatched), marginally centrally stellar-dominated ($1\leq f_{\star}(t_{\rm bar}) < 2$, orange hatched), or while centrally DM-dominated ($f_{\star}(t_{\rm bar}) < 1$, red, sparsely hatched), respectively. While many galaxies are stellar-dominated at the time of bar formation ($76$ per cent), as expected for secular bar formation, $\sim 24$ per cent are DM-dominated. Unlike stellar-dominated galaxies, DM-dominated systems are less susceptible to rapid secular bar formation and likely require external forces to trigger bar growth, as we explore in Section~\ref{sec:tides}.

The right panel of Fig.~\ref{fig:fdm_evolution} shows the redshift evolution of $f_{\star}$ for the galaxies split by $f_{\star}(t_{\rm bar})$ as in the left panel, with the same colours. Discs that never form bars are shown as a thick grey line. Coloured squares mark the median $t_{\rm bar}$ for each population, and the dashed horizontal grey line indicates $f_{\star} = 1$. Looking at the evolution of the galaxies in the different $f_{\star}(t_{\rm bar})$ bins, we see a few key trends: galaxies that form bars when highly stellar dominated (blue line) experience the most rapid growth in $f_{\star}$ at early times, and form their bars on average only $\approx2\,{\rm Gyr}$ after their discs become self-gravitating (half do so by $z\approx1.5$), as expected for secular bar formation \citep{RosasGuevara2022, Frosst2025}. However, as $f_{\star}(t_{\rm bar})$ decreases, bars form later, and $f_{\star}$ grows more slowly and peaks at lower values. The galaxies that form their bars while still DM dominated (red line) tend to show the slowest growth in $f_{\star}$ over time, and are comparable to unbarred discs. Many of these galaxies never become stellar dominated, but when they do it is far after $t_{\rm bar}$, and only at lower redshifts ($z \lesssim 1$). The similarities between the DM-dominated barred galaxies and those that do not form bars suggests that other factors are influencing bar formation. 



\begin{figure*}
	\includegraphics[width=\textwidth]{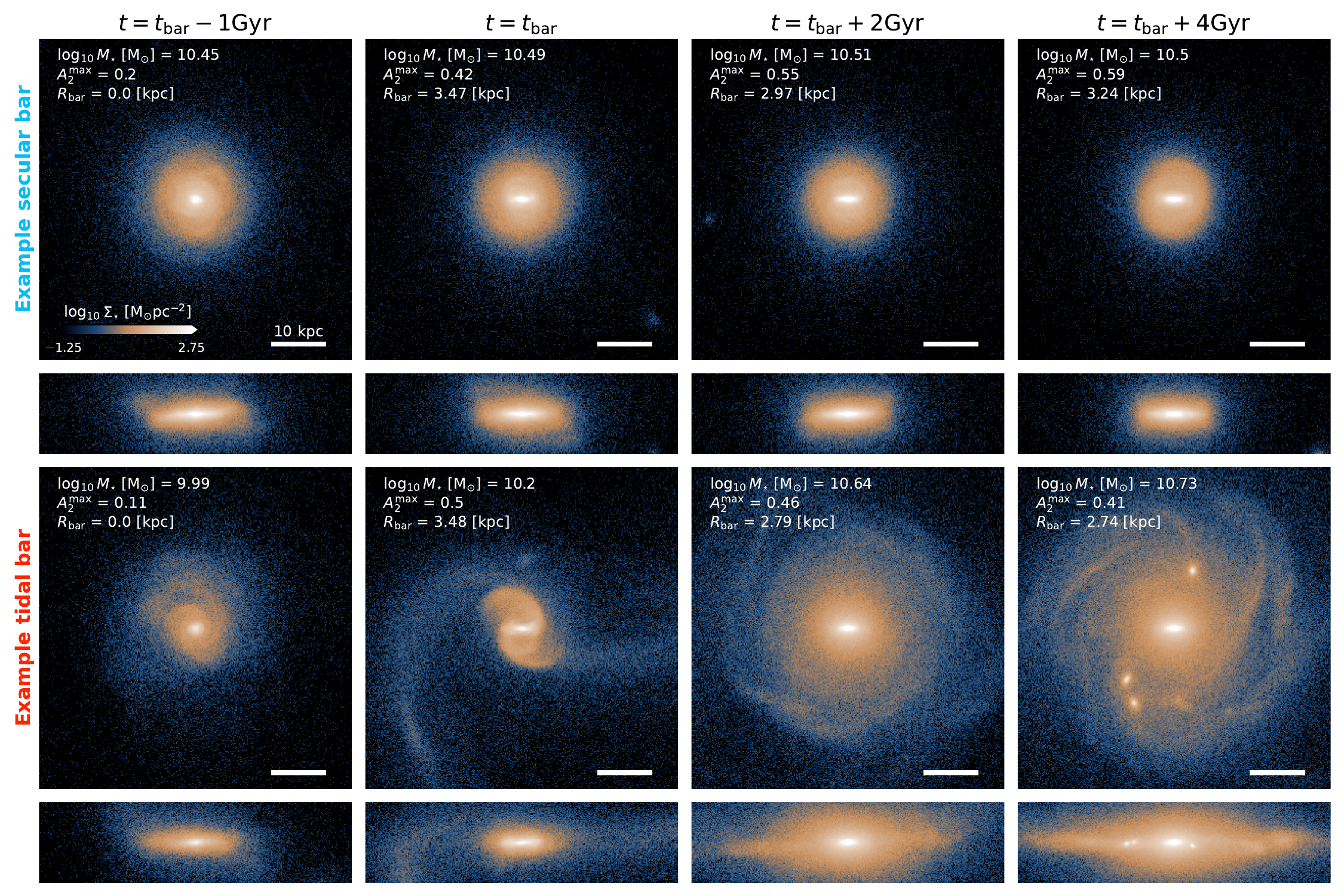}
    \caption{Face-on and edge-on stellar surface mass densities ($\Sigma_{\star}$) of two galaxies that form stellar bars while their discs are either stellar- (top) or DM-dominated (bottom). Their $z=0$ \textsc{Subfind} IDs are 414918 and 388544, respectively. 
    From left to right, the panels show these galaxies 1 Gyr before $t_{\rm bar}$, at $t_{\rm bar}$, and 2 and 4 Gyr after $t_{\rm bar}$, respectively. 
    In the top left corner of face-on panels, we plot the stellar mass, $M_{\star}$ (measured within $r \leq 30$ kpc), the bar strength, $A_{\rm 2}^{\rm max}$, and bar length, $R_{\rm bar}$.  
    Projections are coloured according to the stellar surface mass density. 
    White lines in the bottom right of each panel illustrate the size of $10\,{\rm kpc}$. 
    }
    \label{fig:projection}
\end{figure*}

\begin{figure*}
	\includegraphics[width=\textwidth]{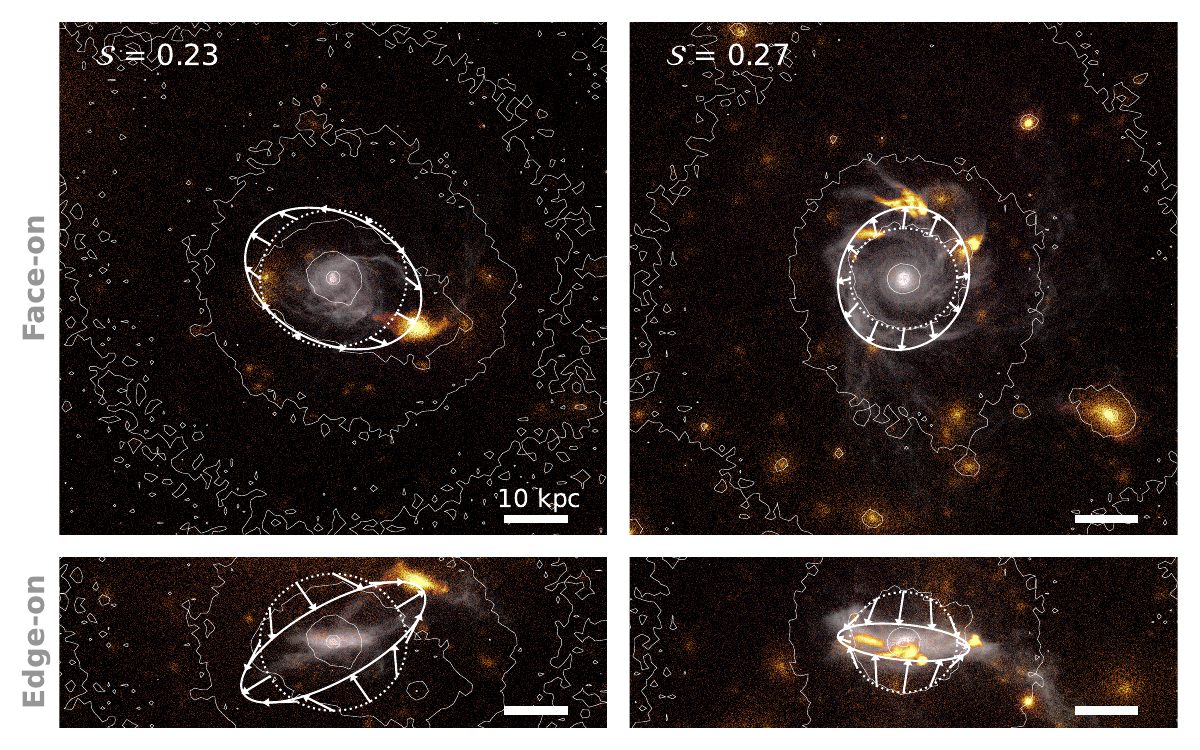}
    \caption{Face-on (top panels) and edge-on (bottom panels) surface mass densities of two example galaxies at $z=0.76$, when $\mathcal{S}$ reaches a maximum prior to the formation of a stellar bar (the $z=0$ \textsc{Subfind} IDs are 441709 and 432106 for the galaxies in the left- and right-hand panel, respectively). The grey colour map corresponds to baryonic particles and cells (stars and gas) bound to the central galaxy whereas orange coloured points show those bound to satellite galaxies, but including the DM. White contours show the total DM density distribution. The dotted white circle in each panel indicates the radius $r_{\rm t} = 4\, r_{\star, 1/2}$ at which we measure the tidal field, $\mathcal{S}$. The white ellipse and associated arrows show how these circles are distorted due to each galaxy's asymmetric tidal field (the magnitude of $\mathcal{S}$ is provided in the upper-left corner of each panel). While satellite galaxies typically dominate the local tidal field, their collective influence must be considered to fully characterise the tidal perturbations experienced by galactic discs.}
    \label{fig:projection_expanded}
\end{figure*}

\subsection{Visual illustration of two TNG50 bars} \label{sec:case_study}
Fig.~\ref{fig:projection} shows the evolution of the face-on and edge-on stellar surface mass density projections of two example galaxies (time moving from left to right). In the top row, the bar forms when stellar dominated ($f_{\star}(t_{\rm bar}) \approx 3.2$), while in the bottom row the bar formed when DM-dominated ($f_{\star}(t_{\rm bar}) \approx 0.1$). While both example galaxies form their bars rapidly, it is clear that they do so for different reasons. The stellar-dominated galaxy forms its bar in isolation, suggesting a secular origin. Conversely, the DM-dominated galaxy in the bottom row forms its bar during the close passage of a nearby satellite, characteristic of tidally induced bar formation. 

Note also that the bars themselves are not easily distinguishable, with roughly similar strengths and lengths, though the secular bar is longer and stronger than the tidal bar after $4\,{\rm Gyr}$. However, the discs of these two example galaxies \textit{are} visually distinct. For instance, after $t_{\rm bar}$, the galaxy that forms the secular bar is thinner than galaxy that forms the tidal bar, which is visually thicker and displays warps, spiral arms, and other non-axisymmetric extended stellar features, including diffuse light. Clearly, bars form in both stellar- and DM-dominated disc galaxies, and these systems may form bars through different formation mechanisms. To proceed, we require a reliable method to characterise the tidal fields acting on bar forming galaxies. 

\begin{figure}
	\includegraphics[width=\columnwidth]{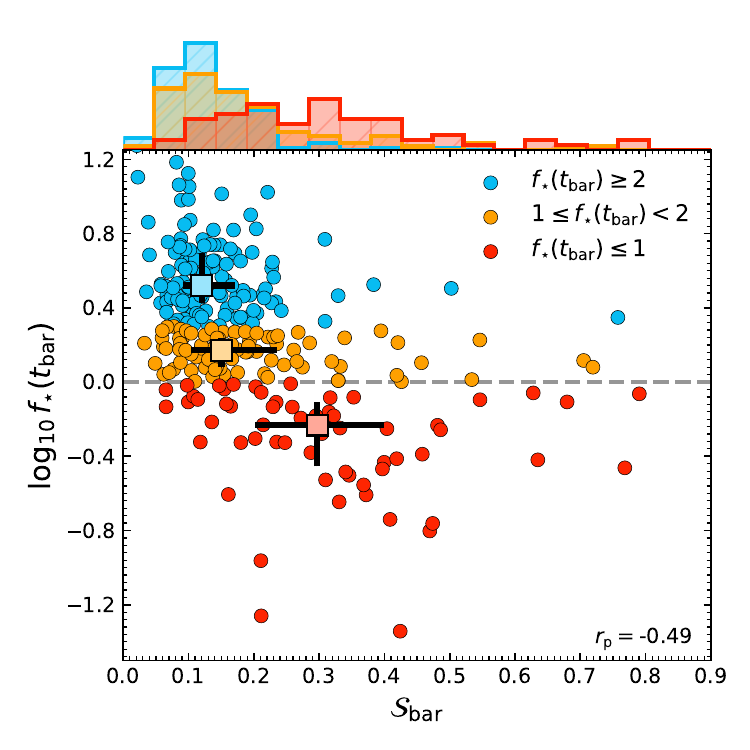}
    \caption{The relationship between central stellar-to-DM mass ratio at the time of bar formation, $f_{\star}(t_{\rm bar})$, and $\mathcal{S}_{\rm bar}$ for galaxies with 1 bar episode. Points are coloured according to $f_{\star}(t_{\rm bar})$, as in Fig.~\ref{fig:fdm_evolution}. The light-coloured squares in the main panel shows the median $f_{\star}(t_{\rm bar})$ and median $\mathcal{S}_{\rm bar}$ for each sub-sample, while the black errorbars show the IQR. The horizontal dashed grey line indicates where galaxies are centrally stellar-dominated (above), or DM-dominated (below). Distributions of $\mathcal{S}_{\rm bar}$ for $f_{\star}(t_{\rm bar})$ sub-sample are shown on the upper x-axis.}
    \label{fig:Ta_MsMh_distribution}
\end{figure}

\section{Tidal fields near bar forming galaxies}\label{sec:tides}
To some degree, all galaxies constantly evolve within external tidal fields that can influence a range of structural features, including disc warps, flares, and stellar bars \citep[e.g.][]{Moore1998, Gnedin2003, RomanoDiaz2008, Oh2008, Pettitt2018}. These tidal asymmetries are often attributed to nearby satellite galaxies and are commonly modelled by treating satellites as point-mass perturbers acting on an otherwise axisymmetric disc potential. This approach has been widely applied in both observational \citep[e.g.][]{Elmegreen1991, Karachentsev1999, Weisz2011, Pearson2016} and simulation-based studies \citep[e.g.][]{Berentzen2004, Lokas2016, Semczuk2017, Peschken2019, Ansar2023, Lopez2024}. However, such metrics over-simplify the complex environments around galaxies and treat satellites independently, often assuming the tidal field is associated with a single satellite that contributes the largest tidal perturbation. In reality, tidal fields arise from complex distributions of stars, gas, and DM, which can interact to either amplify or suppress tidal effects associated with a single passing satellite. Motivated by this, we construct a general estimator of the total tidal field acting on galaxies due to the full surrounding matter distribution.

\subsection{Measuring the local tidal strength}
We quantify the tidal field at a characteristic radius $r_{\rm t}$ (here chosen to be $r_{\rm t} \equiv 4r_{\star,1/2}$)\footnote{This value of $r_{\rm t}$ is suitable for our purposes because it extends beyond the stellar disc; reasonable choices for $r_{\rm t}$ do not strongly influence our results.}, which approximates the size of the galaxy and encloses a total mass $M_{\rm t} \equiv M_{\rm tot}(r\leq r_{\rm t})$, including stars, gas, black holes, and DM. Upon normalising the tidal acceleration by the self-gravity of the material inside $r_{\rm t}$ (i.e., the galaxy) and retaining only the leading-order (quadrupole) terms, the tidal acceleration experienced on a spherical surface of radius $r_{\rm t}$ due to a single point-mass perturber of mass $m_{\rm p}$ at position $\mathbf{r}_{\rm p}$ is given by
\begin{equation} \label{eq:simpleT}
    a_{\rm t}(\hat{n}) = \frac{m_{\rm p} \, r_{\rm t}^3}{M_{\rm t}\, r_{\rm p}^3}\left[ 3 \frac{\mathbf{r}_{\rm p} \otimes \mathbf{r}_{\rm p}}{r_{\rm p}^2} - \mathcal{I}\right]\hat{n} \equiv \mathcal{T}\hat{n},
\end{equation}
where $G$ is the gravitational constant, $\mathcal{I}$ is the identity matrix, $r_{\rm p} = |\mathbf{r}_{\rm p}|$, $\hat{n}$ is the normal vector on the sphere, and $\mathbf{r}_{\rm p}\otimes \mathbf{r}_{\rm p}$ is the outer product of $\mathbf{r}_{\rm p}$ with itself. The dimensionless tidal tensor, $\mathcal{T}$, is the product of a dimensionless scalar $\tilde{\mathcal{T}} = \left(m_{\rm p}\, r_{\rm t}^3\right) / \left(M_{\rm t}\, r_{\rm p}^3\right)$ and a trace-free symmetric matrix with eigenvalues $(+2, -1, -1)$. The scalar $\tilde{\mathcal{T}}$ is commonly referred to as the ``scaled tidal index'' (e.g. \citealp{Ansar2023, Lopez2024}; \citealp[and similar to][]{Elmegreen1991} \citealp[or][]{Karachentsev1999}), and represents the normalised tidal acceleration relative to the self-gravity of the galaxy at the radius $r_{\rm t}$ exerted by a single a point-mass perturber. 

This formulation can be generalised to multiple perturbers, such as a galaxy’s satellite population or, in our case, the total density field beyond $r_{\rm t}$. The total tidal tensor is then given by:
\begin{equation}\label{eq:sumT}
    \mathcal{T} = \frac{r_{\rm t}^3}{M_{\rm t}}\sum_{k, r_{k} > r_{\rm t}}\frac{m_{k}}{r_{k}^3} \left[3\, \frac{\mathbf{r}_k \otimes \mathbf{r}_k}{r_k^2} - \mathcal{I} \right],
\end{equation}
where $m_k$ and $\mathbf{r}_k$ are the masses and positions of particles and fluid elements relative to the galaxy. Unlike in the single perturber case, the eigenvalues of $\mathcal{T}$ in this general form depend on the full spatial distribution of the surrounding matter, but the tensor remains trace-free.

We characterise the strength of the tidal field by the Euclidean norm of the eigenvalues of $\mathcal{T}$, 
\begin{equation}\label{eq:sumTnorm}
    \mathcal{S} = \left(\sum_{j=1}^3 \lambda_{j}^2 \right)^{1/2} = {\rm Tr}\left(\mathcal{T}^2\right)^{1/2}.
\end{equation}
For a single point mass perturber, $\mathcal{S}$ differs from $\tilde{\mathcal{T}}$ by a factor of $\sqrt{1/6}$. We choose to use this 3D scalar tidal strength $\mathcal{S}$, rather than a 2D projection onto the disc plane, because our tests showed that $\mathcal{S}$ more effectively captures relevant features such as symmetric vertical compressions that may influence bar formation\footnote{We provide a simple Python script to measure $\mathcal{T}$ and $\mathcal{S}$ from simulation data at \url{https://github.com/mattfrosst/tidalfields/}.}.

Fig.~\ref{fig:projection_expanded} shows the face-on and edge-on surface mass densities of two example galaxies at $z = 0.76$. Gray denotes all baryonic particles and cells bound to the central galaxy, while orange coloured regions indicate all material associated with satellite galaxies (including DM). Both systems are centrally DM-dominated yet develop stellar bars within one halo dynamical time, $t_{\rm dyn}$, of when they are shown. We define $t_{\rm dyn}$ as the halo free-fall time, 
\begin{equation}
t_{\rm dyn} = \left(\frac{3\pi}{32G(200\rho_{\rm crit})}\right)^{1/2}.
\end{equation} 
One galaxy (left column) is accompanied by a single massive satellite and several less massive ones; the other (right column) is surrounded by multiple satellites of comparable mass. The dotted white sphere in each panel indicates $r_{\rm t}$. We compute $\mathcal{S}$ by summing over all collisionless particles and gas cells within $r_{\rm t} \leq r_k \leq 400\,\mathrm{kpc}$ of the galaxy centre\footnote{Because $\mathcal{T} \propto r^{-3}$, the choice of outer radius does not affect our results, provided it exceeds a few times $r_{\rm t}$.}, regardless of whether they are gravitationally bound. Despite differing satellite distributions, the two galaxies exhibit similar values of $\mathcal{S}$ (top-left corner of each face-on panel). 

Solid white ellipses and arrows in Fig.~\ref{fig:projection_expanded} illustrate how asymmetric tidal fields from nearby satellites cause accelerations normal to the face of the dotted spheres, with each ellipse having the same volume as its corresponding sphere. In the left column, the tidal field clearly points toward the dominant satellite and compresses it orthogonally. As this field is dominated by a single satellite, $\mathcal{T}$ has only one positive eigenvalue. In the right column, the tidal field is more complex: it is stretched in the plane of the disc toward a triplet of satellites (just above the dotted circle in the top-right panel) and compressed perpendicularly to them. In this case, $\mathcal{T}$ has \textit{two} positive eigenvalues, a situation not possible in the point-mass assumption. This shows that $\mathcal{S}$ can flexibly capture the deformation of the tidal field associated with a variety of environments. 



\subsection{Secular versus tidal bars}
We quantify the local tidal field during bar formation using $\mathcal{S}_{\rm bar}$, defined as the maximum $\mathcal{S}$ within one $t_{\rm dyn}$ before $t_{\rm bar}$; larger values correspond to stronger tidal forces just before bar formation. The main panel of Fig.~\ref{fig:Ta_MsMh_distribution} plots the relationship between $\mathcal{S}_{\rm bar}$ and $f_{\star}(t_{\rm bar})$. The light-coloured squares show that the median value of $\mathcal{S}_{\rm bar}$ increases as the median value of $f_{\star}(t_{\rm bar})$ declines (with a Spearman rank correlation coefficient of $r_{\rm p} = -0.49$). This suggests that while stellar-dominated discs can form bars secularly under relatively weak tidal perturbations (low $\mathcal{S}_{\rm bar}$), increasingly DM-dominated discs require stronger perturbations, and often experience large tidal interactions around the time their bars first form (high $\mathcal{S}_{\rm bar}$). 

This is further corroborated by comparing the distributions of $\mathcal{S}_{\rm bar}$ from the most highly stellar- and DM-dominated populations on the top axis (blue and red PDFs). These distributions highlight the diversity and of tidal forces bar forming galaxies experience. The continuous, unimodal, but overlapping distributions of $\mathcal{S}_{\rm bar}$ for different $f_{\rm bar}(t_{\rm bar})$ bins suggests that secular and tidal bars are not always separable, stressing that any threshold between the two is meaningful only in controlled experiments\footnote{Others have attempted to justify other, comparable methods to define tidal and secular bars in cosmological simulations. We show in Appendix~\ref{apx:altmethods} that arbitrary thresholds in $\mathcal{S}_{\rm bar}$ can be roughly consistent with these methods.}. Ultimately, the most stellar- and DM-dominated distributions are found to be statistically distinct, further suggesting that bars with different $f_{\rm bar}(t_{\rm bar})$ are not forming under the same processes. Thus, having a stellar-dominant disc is not a prerequisite for bar formation generally, but \textit{is} required for secular bar formation; DM-dominated discs tend to form bars through tides. 


\begin{figure}
	\includegraphics[width=\columnwidth]{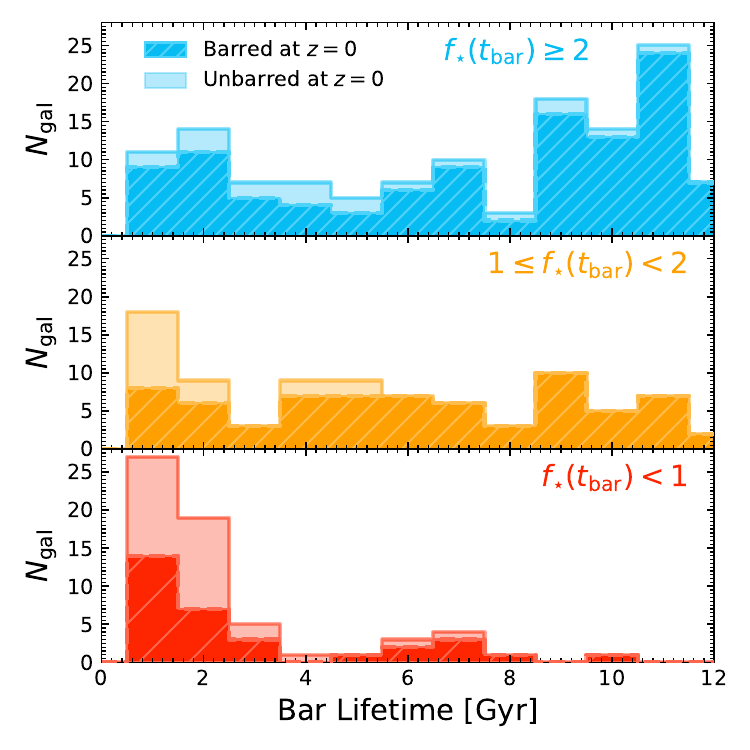}
    \caption{The distribution of bar lifetimes for galaxies in each $f_{\star}(t_{\rm bar})$ bin, with $f_{\star}(t_{\rm bar})$ decreasing from top-to-bottom. Hatched histograms with dashed outlines show the distribution for galaxies that are still barred at $z=0$, which should be treated as a lower limit. The open histograms with solid outlines show the distribution of galaxies that are unbarred at $z=0$; these are added on top of the hatched histograms. }
    \label{fig:Barage_MsMh_distribution}
\end{figure}

Note the significant scatter along the relation, which suggests that some stellar-dominated discs form bars after experiencing large tidal forces. This is expected, as a high $f_{\star}(t_{\rm bar})$ does not preclude a galaxy from experiencing a strong interaction or merger. Furthermore, other factors like the disc properties \citep[e.g.][]{Ghosh2022} or the orbital configuration of the tidal interaction \citep[e.g.][]{Lokas2018, Peschken2019} can influence bar formation, increasing the scatter in this plane. The Milky way provides a possible example: its bar is thought to have formed after a merger, and may have been long-lived \citep[e.g.][]{Grady2020, Sanders2024}. This scenario is consistent with the range of behaviours observed in the stellar-dominated discs of our sample.

In Fig.~\ref{fig:Barage_MsMh_distribution}, we show the distribution of bar lifetimes for galaxies with different $f_{\star}(t_{\rm bar})$. The distributions are hatched for galaxies that are still barred at $z=0$, in which case the lifetimes are considered lower limits, and open if the bar no longer exists at $z=0$. Bars in stellar-dominated discs are typically long-lived (median $\approx9,{\rm Gyr}$) and persist to $z=0$, whereas bars in DM-dominated discs are usually short-lived (median $\approx2,{\rm Gyr}$) and disappear by $z=0$. Indeed, the majority of bars that form with $\mathcal{S}_{\rm bar} \geq 0.2$ last for less than $\lesssim 2\,{\rm Gyr}$ (most of these are DM-dominated, and they all experienced a merger within $\pm 2t_{\rm dyn}$ of $t_{\rm bar}$, see Appendix~\ref{apx:altmethods}). Note, however, that bars with high $\mathcal{S}_{\rm bar}$ can become long-lived features if they form in galaxies had high $f_{\star}(t_{\rm bar})$; this suggests that the bars lifetime is set more by the disc's internal properties than the bar formation mechanism. This agrees with previous simulations showing that secular bars formed in isolation tend to be long-lived \citep[e.g.][]{Athanassoula2013}, whereas bars formed tidally or in cosmological settings are often short-lived \citep[][]{RomanoDiaz2008, Peschken2019, Ansar2023}. 

Why do some bars disappear by $z=0$? We find that roughly half of all bars that are terminated do so after a period of high $\mathcal{S}$, frequently associated with a merger, while the rest display a gradual, secular dampening \citep[with some becoming bulge-dominated systems, see][]{Gargiulo2022}. Note however that not all mergers lead to the loss of a bar \citep[e.g.][]{Martinez2017, Lokas2018, Zana2018a}. Understanding the resilience of bars to some mergers and interactions, but not others, is beyond the scope of this work. 

\section{The properties of secular and tidal bars} \label{sec:results}

As shown so far, in our \textsc{TNG50} sample, bars form in both stellar- and DM-dominated discs, mainly through secular evolution and tidal interactions, respectively. We now examine whether other disc properties influence bar formation, whether they produce bars with different characteristics, and whether tidal and secular bars can be distinguished in low-$z$ galaxies using observables.

\subsection{The properties of tidal and secular bars}
\begin{figure}
	\includegraphics[width=\columnwidth]{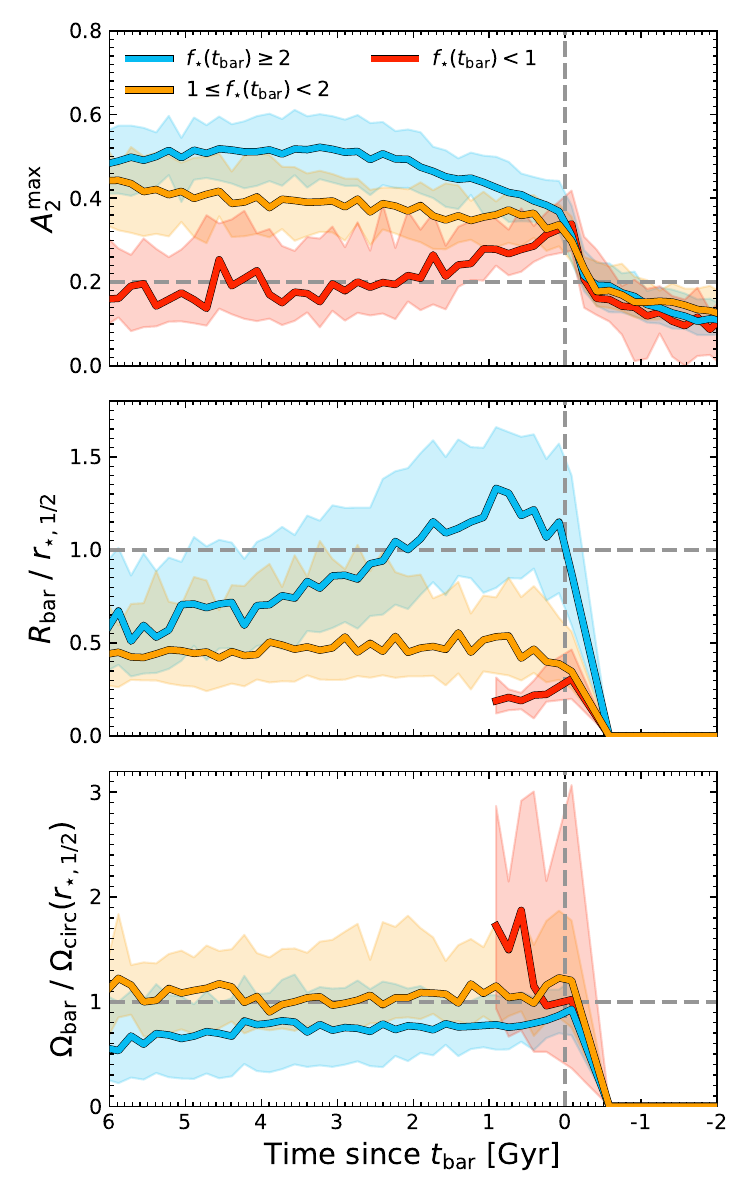}
    \caption{From top to bottom, we show the bar strength, length, and pattern speed, respectively, as a function of lookback time normalised by the bar formation time, $t_{\rm bar}$.
    Galaxies that form bars under different $f_{\star}(t_{\rm bar})$ are coloured as in Fig.~\ref{fig:fdm_evolution}. 
    Shaded regions indicate the IQR of each population. 
    In the upper panels, we show as a dashed grey line the $A_{2}^{\rm max} = 0.2$ threshold used to identify bars. 
    }
    \label{fig:bar_props_evo}
\end{figure}

In Fig.~\ref{fig:bar_props_evo}, from top to bottom we show the evolution of the bar strength ($A_2^{\rm max}$), bar length ($R_{\rm bar}$, normalised by $r_{\star, 1/2}$), and bar pattern speed ($\Omega_{\rm bar}$, normalised by $\Omega_{\rm circ}$, the disc pattern speed at $r_{\star,1/2}$, where $\Omega_{\rm circ} = V_{\rm circ}(r_{\star,1/2})~r_{\star,1/2}^{-1}$, respectively). The median trends for bars with different $f_{\star}(t_{\rm bar})$ are coloured as before, with the blue, orange, and red lines indicating galaxies that formed their bars when they were highly stellar-dominated, marginally stellar-dominated, and DM-dominated, respectively. From the results of Fig.~\ref{fig:Ta_MsMh_distribution}, we have also seen that these populations become increasingly tidally induced as they become less stellar-dominated. To enable comparisons between bars at similar stages of their evolution, time is defined such that $t=0$ marks the onset of the bar episode. In the lower two panels, galaxies are only included in the bin median where their bar exists, and there are more than 20 bars available for measurement at that time; conversely, we show $A_{2}^{\rm max}$ regardless of whether galaxies are actively barred. In Appendix~\ref{apx:bar_props}, we show the un-normalised bar properties as a function of $M_{\star}$ at the time of bar formation. 

The top panel of Fig.~\ref{fig:bar_props_evo} shows that the bar strength evolves differently for galaxies with different $f_{\star}(t_{\rm bar})$. At the time of bar formation, all populations show comparable $A_{2}^{\rm max}$, but quickly diverge thereafter. Galaxies that were more stellar-dominated at $t_{\rm bar}$ produce bars that more rapidly grow stronger over time. In contrast, bars in DM-dominated discs show a gradual decline in $A_{2}^{\rm max}$ after $t_{\rm bar}$, in agreement with the short bar lifetimes found for these galaxies in Fig.~\ref{fig:Barage_MsMh_distribution}. These results reiterate that bars that form after high tidal forces are weaker and grow more slowly -- secular bars grow to be stronger than tidal bars, given time. 


\begin{figure*}
	\includegraphics[width=\textwidth]{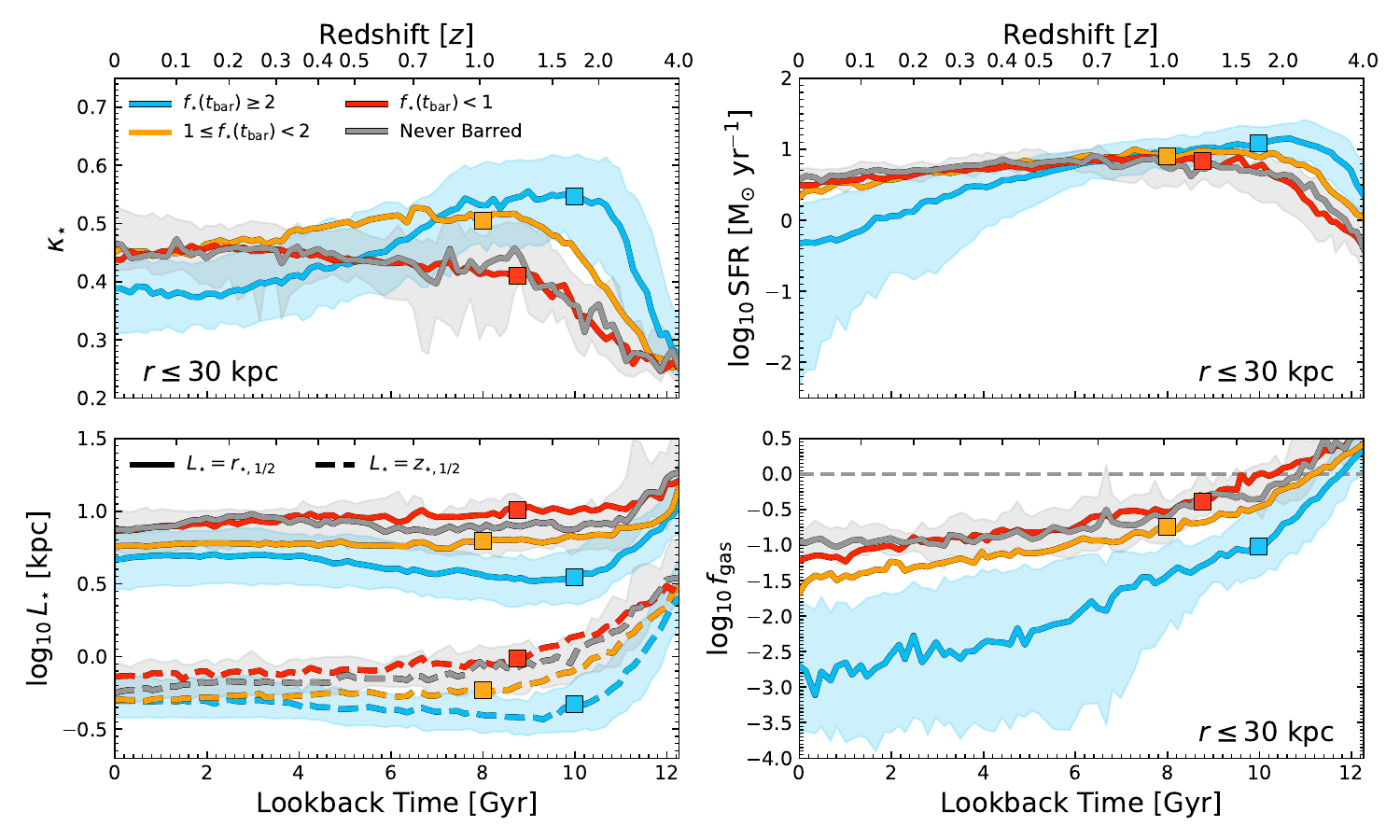}
    \caption{We show the time evolution of various galaxy morphology indicators for galaxies that form bars under different $f_{\star}(t_{\rm bar})$, coloured as in Fig.~\ref{fig:fdm_evolution}. 
    We show the fraction of energy invested in co-rotation, $\kappa_{\star}$ (within $r\leq30\,{\rm kpc}$; upper left), star formation rate (measured over the past $50\,{\rm Myr}$ and within $r\leq30\,{\rm kpc}$; upper right), the stellar half mass radius and height (solid and dashed lines, respectively; lower left), and the gas-to-stellar mass ratio, $f_{\rm gas}$ (measured within $r\leq 30\,{\rm kpc}$; lower right). 
    Coloured squares indicate the median bar formation time for these populations. 
    Shaded areas show the IQR of select sub-samples.}
    \label{fig:disc_props}
\end{figure*}

The middle panel of Fig.~\ref{fig:bar_props_evo} reveals further contrasts in the evolution of the bar length relative to disc's $r_{\star,1/2}$. During the secular formation of bars in stellar-dominated discs, the bar length grows significantly relative to the galaxies size. Likewise, during the tidal interactions that predominantly lead to the formation of bars in DM-dominant discs, the bar exhibits a rapid increase in $R_{\rm bar}$ (at $t_{\rm bar}$, many of the longest bars in our sample are tidal bars). However, the discs that form tidal bars tend to be far more extended than those that form secular bars, resulting in tidal bars that are comparatively shorter than secular bars relative to the disc size (shown in Section~\ref{sec:disc_props}). 
As a result, the relative lengths of tidal bars in DM-dominated discs are often shorter than the secular bars in stellar-dominated discs; if the bar can persist for $\approx4\,{\rm Gyr}$, we see that this effect is reduced, and all surviving bars tend towards $R_{\rm bar}/r_{\star,1/2}\approx0.5$. 
After this the normalised lengths of bars do not evolve significantly, suggesting that, once embedded and stable, bars grow simultaneously with their discs \citep[e.g.][]{Kim2021} regardless of formation mechanism or $f_{\star}(t_{\rm bar})$.

Finally, the bottom panel of Fig.~\ref{fig:bar_props_evo} shows the evolution of the bar pattern speed is related to $f_{\star}(t_{\rm bar})$: bars in more highly DM-dominant discs reach higher peaks in $\Omega_{\rm bar}/\Omega_{\rm circ}$. Following this peak, if the bars survive, they exhibit a constant pattern speed relative to the disc \citep[but show a decline in the un-normalised $\Omega_{\rm bar}$, consistent with both theoretical expectations and previous simulations;][]{Tremaine1984, Weinberg1985, Zavala2008}. As bars form in progressively more stellar-dominated discs, the relative pattern speed peaks at lower values. The most DM-dominated discs, for instance, show the largest increase in $\Omega_{\rm bar}/\Omega_{\rm circ}$ before their bars are terminated. Again, this is likely related to the systematic differences in $r_{\star,1/2}$ for galaxies with different $f_{\star}(t_{\rm bar})$ (though we show in Appendix~\ref{apx:bar_props} that bars formed in DM-dominated discs have physically slower $\Omega_{\rm bar}$ than more stellar-dominated bar forming galaxies). Note however, that these trends are mostly $\mathcal{S}_{\rm bar}$ agnostic: at fixed $S_{\rm bar}$, pattern speeds evolve based on their $f_{\star}(t_{\rm bar})$. These results echo those of \citet{Zheng2025}, who found that tidally induced bars that were otherwise stable to secular bar formation rotated more slowly than those which were secularly unstable \citep[see also][]{Martinez2017}. This differs slightly from the predictions of \citet{Miwa1998} who suggest that all tidal bars should show lower $\Omega_{\rm bar}$ than secular bars \citep[e.g.,][]{Lokas2014, Lokas2018, Pettitt2018}. It seems that differences in $\Omega_{\rm bar}$ are driven more by the differences in the internal properties of galaxies, particularly $f_{\star}$, than by the different formation mechanisms.

Ultimately, while secular and tidal bars show distinct trends in strength, length, and pattern speed, these differences become increasingly subtle with time, and are further complicated by differences in galaxy properties. Once a bar is well-established, its properties alone are likely insufficient to unambiguously determine its formation mechanism. Thus, early evolution can hint at a bar’s origin, especially when paired with contextual information about the host galaxy or the nearby environment, but the bars themselves do not always retain a clear imprint of their formation history.

\begin{figure*}
	\includegraphics[width=\textwidth]{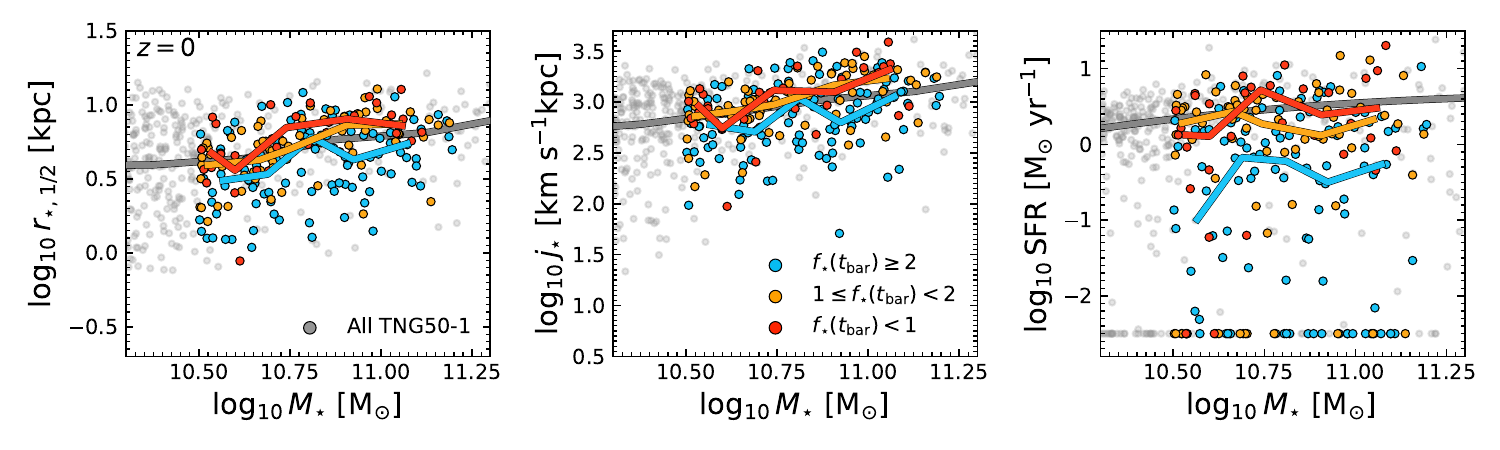}
    \caption{We show three scaling relations for our sample at $z=0$ and $z=1$ in the top and bottom rows, respectively. From left to right, we show the size-mass relation, Fall relation, and star formation main sequence (SFMS). Galaxies are coloured based on $f_{\star}(t_{\rm bar})$, as in Fig.~\ref{fig:fdm_evolution}, but are only coloured if they are currently barred at these redshifts. For comparison, all \textsc{TNG50} galaxies are included as light grey points. The medians of each population are shown as thick lines of the same colours. }
    \label{fig:scaling_relations}
\end{figure*}

\subsection{Galaxy properties} \label{sec:disc_props}
While our analysis confirms that strongly self-gravitating disc galaxies are particularly susceptible to secular bar formation \citep[see also][]{ELN1982, BlandHawthorn2023, Frosst2024}, other properties like disc thickness \citep[e.g.][]{Klypin2009, Aumer2017, Ghosh2022} or gas fraction \citep[e.g.][]{Berentzen2004, Bournaud2005} may also be important. 

To investigate the link between galaxy structure and bar formation, in Fig.~\ref{fig:disc_props} we present the redshift evolution of four key quantities that trace a galaxy’s structure and susceptibility to secular bar formation. 
A striking feature is the link between $f_{\star}(t_{\rm bar})$ and the assembly of galaxy stellar structure. While it is well understood that barred galaxies assemble their mass faster than unbarred galaxies (\citealp[found in \textsc{TNG}, e.g.,][]{RosasGuevara2022, Khoperskov2024}, \citealp[but also recent observations][]{FraserMcKelvie2020a}), we see now that galaxies that are more stellar-dominated at $t_{\rm bar}$ assemble their discs earlier than galaxies that are more DM-dominated at $t_{\rm bar}$. This is indicated by higher $\kappa_{\star}$ at high $z$, but also by how these galaxies become thinner and more compact at earlier times. In contrast, the galaxies that form bars in DM-dominated discs tend to assemble their discs later, and, even by $z=0$, remain less compact than galaxies that formed bars with higher $f_{\star}(t_{\rm bar})$. These structural trends are consistent with the idea that secular bars form predominantly in thin, kinematically cold discs prone to internal instabilities, whereas tidal bars form in more stable discs that require external perturbations to induce bar formation \citep[e.g.][]{Aumer2017, Klypin2009, Fragkoudi2017, Ghosh2022}. 

Interestingly, while the most highly stellar dominated galaxies assemble their discs earliest, and thus form their bars earliest, by $z=0$ they have the lowest $\kappa_{\star}$ of all sub-samples. We speculate that this is related to the presence of the long-lived bars found in these galaxies, which should increase the central velocity dispersion as they evolve \citep[e.g.][]{Das2008}. \citep{Zana2022} also suggests that the presence of bars may lower $\kappa_{\star}$. It is possible that some bars in MW-mass disc galaxies are missed due to our $z=0$ selection on this parameter. 


Another notable (but related) distinction emerges in the SFR and gas content of these galaxies. Galaxies that go on to form bars in highly stellar-dominated discs are significantly more gas-poor than their DM-dominated counterparts, even at early times\footnote{Note that, while there is no statistical difference in $M_{\star}$ between the different $f_{\star}(t_{\rm bar})$ bins at $z=0$, stellar-dominated discs do assemble their stellar mass earlier than DM-dominated discs, which may influence these results.}. They also tend to be more highly star forming at high redshifts, but more quenched by $z=0$ -- as a consequence, they are often more stellar in-situ dominated, having assembled most of their mass through star formation. As $f_{\star}(t_{\rm bar})$ gets lower, the galaxies retain higher $f_{\rm gas}$ values over time, and exhibit flatter star formation histories; more of their mass is assembled via mergers. This may be consistent with previous work showing that gas-rich discs are less prone to secular bar formation \citep[e.g.,][]{Athanassoula2013b}, although recent work has suggested that $f_{\rm gas}$ makes little difference at fixed $f_{\star}$ \citep{Bland-Hawthorn2024}. It has also been shown that bars in gas-rich galaxies tend to be more shortly-lived \citep[e.g.,][]{Berentzen2004, Bournaud2005, Bland-Hawthorn2024}, in agreement with our results in Fig.~\ref{fig:Barage_MsMh_distribution}. It is possible that the combination of high gas fractions and external perturbations in DM-dominated tidal bar hosts may weaken or ultimately destroy nascent bars through central mass concentration and induced turbulence \citep[e.g.,][]{Friedli1993, Bournaud2005, Peschken2019}. Ultimately, caution is warranted when interpreting trends with $f_{\rm gas}$, as they may be influenced by strong SMBH feedback in the \textsc{TNG} model, often associated with galaxies that form secular bars in this mass range, potentially complicating direct causal interpretations \citep{Frosst2025}.

Finally, galaxies that never form bars are on average the most extended, thickest, most centrally DM dominated, highly star forming, gas-rich galaxies in our entire sample at all times \citep[see also][]{Fragkoudi2020, Lu2024}. These galaxies tend to form their discs later than the galaxies that form bars, though by $z=0$, they tend to have the largest fraction of stellar mass in ``disc-like'' orbits, with high $\kappa_{\star}$. 
Interestingly however, the properties of unbarred galaxies are usually indistinguishable from the properties of DM-dominated bar forming galaxies. Their resistance to bar formation is likely due to a combination of factors: strong central DM-dominance stabilises the disc against tidal perturbations (see Fig.~\ref{fig:fdm_evolution}), and it is possible that these galaxies simply never experience an interaction with the correct properties to trigger bar formation in their otherwise stable discs. 

A possible criticism of Fig.~\ref{fig:disc_props} is that we include all bars regardless of bar lifetimes, as not all galaxies in each bin remain barred until $z=0$ (e.g., Fig.~\ref{fig:Barage_MsMh_distribution}). To address this, in Fig.~\ref{fig:scaling_relations} we show (from left-to-right) the size ($r_{\star,1/2}$)-mass relation, the specific angular momentum ($j_{\star}$)-mass relation, and the star formation main sequence at $z=0$; $j_{\star}$ and SFR measured within $r\leq30\,{\rm kpc}$, with SFR measured over the past $50\,{\rm Myr}$. Now, the galaxies in the $f_{\star}(t_{\rm bar})$ sub-samples are included only if they are actively barred at $z=0$. The median trends for the actively barred galaxies are shown as lines of the same colours. For comparison, all \textsc{TNG50} galaxies are included as grey points, with the light grey line indicating the median. 

As expected from Fig.~\ref{fig:disc_props}, by $z=0$ actively barred galaxies in the most stellar-dominated discs at $t_{\rm bar}$ have lower $r_{\star,1/2}$, $j_{\star}$, and SFR at fixed mass, even when compared to the entire \textsc{TNG50} median. However, the differences between the moderately stellar-dominated, DM-dominated, and actively unbarred populations are less noticeable. Ultimately, while clear separations between the $f_{\star}(t_{\rm bar})$ sub-samples are not possible in these spaces, nearly all actively barred galaxies at $z=0$ with compact $r_{\star,1/2}$, low $j_{\star}$, and significant SFR quenching have $f_{\star}(t_{\rm bar}) \geq 2$. If observed, bars in such low$-z$ galaxies could be tentatively classified as secular. This once again makes it clear that distinguishing between bar formation mechanisms at any given redshift is very difficult, especially without access to individual bar histories and formation times. 

Fianlly, we emphasise that our results are moderated by a number of caveats. Notably, bar formation in \textsc{TNG50} is correlated with the onset of strong SMBH feedback in this mass range \citep{Frosst2025}, which may bias the connections between disc and bar properties, particularly bar lifetimes and $f_{\star}$. Moreover, our analysis focused on a single simulation (\textsc{TNG50}) and a carefully selected sample of galaxies that are discs at $z=0$. This introduces potential biases, particularly as other large-volume simulations with coarser resolution have reported a greater prevalence of tidally induced bars \citep[e.g.][]{Peschken2019} or drastically lower bar fractions generally \citep[e.g.][]{Reddish2022}. In addition, we have studied only one specific feedback model; prior work shows that the nature and frequency of bars can vary with the feedback prescriptions \citep{Zana2019}. Our results therefore assume that \textsc{TNG50} galaxies have sufficient resolution at all times to model both secular and tidally induced bar formation, and that the subgrid physics models are realistic.

\section{Summary and Conclusions}\label{sec:conclusions}
In this work, we studied the evolution of 307 disc galaxies from \textsc{TNG50} with roughly MW-mass at $z=0$ as a function of their central stellar-to-DM mass fraction ($f_{\star}$) at the time of bar formation ($t_{\rm bar}$). Preliminary visual inspection suggested that some of these bars were formed during tidal interactions, rather than secular processes, particularly if their discs were centrally DM-dominated at $t_{\rm bar}$. To quantify this, we introduced a new metric for measuring the tidal fields acting on galaxies, $\mathcal{S}$ (Eq.~\ref{eq:sumTnorm}), which considers not only the tidal forces imparted simultaneously by nearby passing and merging galaxies, but the entire environment surrounding the galaxy. We used this metric to measure the tidal field present prior to bar formation, $\mathcal{S}_{\rm bar}$. Below, we highlight our main results, focusing on galaxies with only one bar episode (nearly $66$ percent of the total sample, see Fig.~\ref{fig:bar_fraction}). 

\begin{enumerate}

    \item Many bars form in centrally stellar-dominated discs ($f_{\star}(t_{\rm bar}) \geq 1$), but $\approx24$ per cent of our sample form bars while DM-dominated (right panel, Fig.~\ref{fig:fdm_evolution}), and thus while their discs are secularly stable against internally-driven bar formation. Furthermore, the galaxies central mass assembly is highly segregated by $f_{\star}(t_{\rm bar})$: as $f_{\star}(t_{\rm bar})$ decreases, bars form later, $f_{\star}$ peaks lower, and the growth rate of $f_{\star}$ decreases (left panel, Fig.~\ref{fig:fdm_evolution}). Interestingly, the galaxies that form bars in the most DM-dominated discs have similar $f_{\star}$ at all times to galaxies that never form bars. 

    \item We found a moderate negative correlation between $f_{\star}(t_{\rm bar})$, and $\mathcal{S}_{\rm bar}$, the tidal force a galaxy experiences prior to bar formation. As galaxies become more centrally DM-dominated and secularly stable against bar formation, they require increasingly large tidal forces to initiate a bar episode (Fig.~\ref{fig:Ta_MsMh_distribution}). As a result, stellar-dominated discs form bars primarily through secular evolution, while DM-dominated discs more frequently form bars after strong tidal interactions, during mergers or flybys. 
    
    \item Furthermore, bars form across a broad, unimodal, continuous distribution of tidal environments (top distributions, Fig.~\ref{fig:Ta_MsMh_distribution}). This indicates that secular and tidal bar formation cannot be arbitrarily separated in a cosmological context. Additionally, bars never form in perfectly isolated conditions (i.e., we never measure $\mathcal{S_{\rm bar}} = 0$) as often assumed in idealised simulations of secular bar formation. 

    \item Secular, centrally stellar-dominant bars are frequently long-lived, while tidal, DM-dominant bars are often transient features (Fig.~\ref{fig:Barage_MsMh_distribution}). However, if a stellar-dominated galaxy forms a bar due to a tidal interaction, it also tends to be long-lived. This indicating that the disc's internal properties are more important for determining bar lifetimes than the bar formation mechanism. 

    \item The properties of bars themselves are not reliable indicators of their formation pathway or $f_{\star}$ history, unless measured within a few Gyr of $t_{\rm bar}$ (Fig.~\ref{fig:bar_props_evo}). First, differences in bar properties seem to be driven more by the internal disc properties, rather than the formation mechanism. Second, while secular / stellar-dominated and tidal / DM-dominated bars might initially show distinct trends in strength, length, and pattern speed, these differences become increasingly subtle with time. Once a bar is well-established, its properties alone are insufficient to unambiguously determine its formation mechanism. 
    
    \item The evolution of discs that form different bars is distinct (Fig.~\ref{fig:disc_props}): galaxies that form secular / stellar-dominated bars exhibit compact, thin, quenched, gas-poor discs which form their discs relatively early ($z \gtrsim 2$). Conversely, galaxies that form tidal / DM-dominated bars -- or never form bars -- tend to be thicker, more extended, more gas-rich and highly star forming, and only become disc-like at later times ($z \lesssim 1$). That the disc properties of tidal / DM-dominated bars and galaxies that never form bars are similar highlights the importance of the tidal interactions for bar formation in DM-dominated discs, and the importance of the internal disc properties for bar formation in stellar-dominated discs. 

    \item Despite the differences in the evolutionary histories, at any fixed redshift it is challenging to distinguish between discs that formed bars through different mechanisms, or while under different $f_{\star}(t_{\rm bar})$, as the individual evolutionary histories significantly complicate the picture (Fig.~\ref{fig:scaling_relations}). However, at $z=0$ the galaxies that formed bars in the most stellar-dominated discs (most of which are secularly forming) do have smaller sizes, lower specific angular momentum, and lower star formation rates, on average. 

\end{enumerate}


Looking ahead, future studies should test the robustness of these findings across cosmological simulations with varying resolutions and feedback models. Higher output cadence and multi-simulation comparisons will be valuable for disentangling the causality between strength of tidal events, the onset of bar formation, and the assembly of disc structure. Observationally, upcoming integral field unit (IFU) surveys of disc galaxies (particularly those capable of resolving stellar kinematics, e.g., \textsc{TIMER}; \citealt{Gadotti2019}, \textsc{PHANGS-MUSE}; \citealt{Emsellem2022}, and \textsc{GECKOS}; \citealt{vandeSande2023, FraserMcKelvie2024}, among others) may be able to constrain bar origins through stellar population gradients \citep[see][]{Molaeinezhad2017}, bar age estimates \citep[e.g.][]{desaFreitas2023, deSaFreitas2025}, or other structural diagnostics. Ultimately, understanding how, when, and why bars form will provide crucial insight into the role they play in galaxy formation and evolution in a cosmological context.


\section*{Acknowledgements}
We thank Elizabeth Iles for helpful conversations. ADL and DO acknowledge financial support from the Australian Research Council through their Future Fellowship scheme (project numbers FT160100250, FT190100083, respectively). AFM gratefully acknowledges the sponsorship provided by the European Southern Observatory through a research fellowship. This research was undertaken with the assistance of resources and services from the National Computational Infrastructure (NCI), which is supported by the Australian Government. The IllustrisTNG simulations were undertaken with compute time awarded by the Gauss Centre for Supercomputing (GCS) under GCS Large-Scale Projects GCS-ILLU and GCS-DWAR on the GCS share of the supercomputer Hazel Hen at the High Performance Computing Center Stuttgart (HLRS), as well as on the machines of the Max Planck Computing and Data Facility (MPCDF) in Garching, Germany. This work has benefited from the following public \textsc{Python} packages: \textsc{Scipy} \citep{Virtanen2020}, \textsc{Numpy} \citep{Harris2020}, and \textsc{Matplotlib} \citep{Hunter2007}. 

\section*{Data Availability}

The \textsc{TNG} simulation outputs used for our analysis are publicly available at \url{http://www.tng-project.org}; see \citet{Nelson2019a}, \citet{Pillepich2018a} for further information. The \textsc{TNG50} MW/M31-like sample is also publicly available at the above link, but \citet{Pillepich2024} for details. A simple Python script to calculate the total tidal tensor, $\mathcal{T}$, or the tidal field strength, $\mathcal{S}$, can be found at \url{https://github.com/mattfrosst/tidalfields/}. Additional data or analysis codes can be made available upon reasonable request. 
 


\bibliographystyle{mnras}
\bibliography{references} 


\appendix
\section{Measuring bar properties}\label{apx:bar_examples}
\begin{figure*}
	\includegraphics[width=\textwidth]{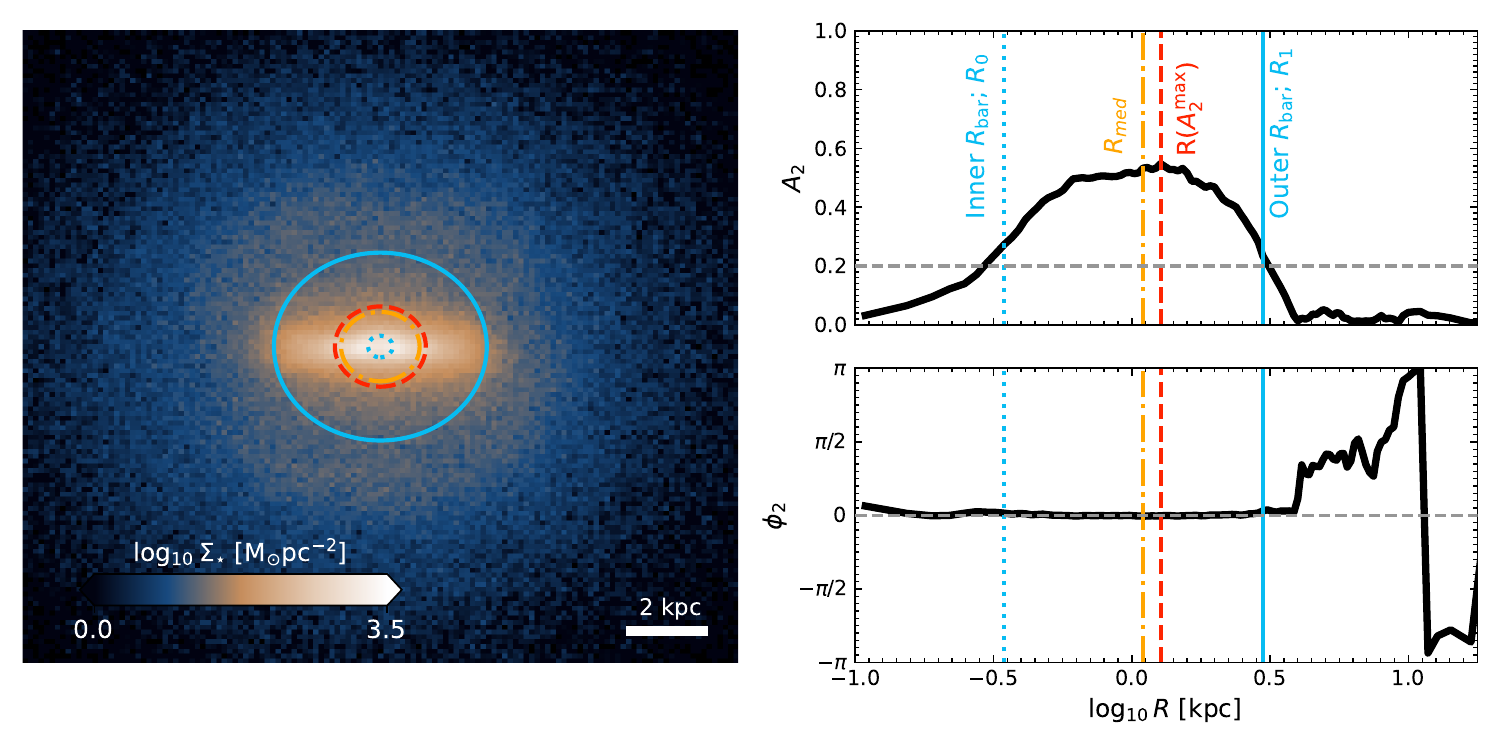}
    \caption{The face-on stellar surface mass density (left; as in Fig.~\ref{fig:projection}), and radial $A_{2}(R)$ and $\phi_{2}(R)$ profiles, (right, top and bottom, respectively). 
    Light blue dotted and solid lines indicate the inner and outer bar radii, $R_{0}$ and $R_{1} = R_{\rm bar}$, respectively. 
    The orange dashed-dotted line shows $R_{\rm med}$, the median radius between $R_{0}$ and $R_{1}$. 
    The dashed red line shows the radius of $A_{2}^{\rm max}$. }
    \label{fig:example_bars}
\end{figure*}
Much of our analysis relies on the accurate measurement of bar properties. We have established that our measurements use a Fourier decomposition to determine $A_{2}$ (Eq.~\ref{eq:A2}) and $\phi_{2}$ (Eq.~\ref{eq:phi2}). These properties are measured as a function of radius, calculated on all stellar particles in cylindrical radial bins centred on the galaxies centre of potential (measured from all bound particles, and, in practice, at the centre of the DM halo). We include exactly $N_{\rm part} = 10^4$ disc particles per radial bin to keep the Poisson noise, scaling as $\sigma_{A_{2}} = 1 / \sqrt{N_{\rm part}}$, at $1$ per cent while maintaining good radial resolution at our stellar mass range in \textsc{TNG50} -- higher $N_{\rm part}$ reduces noise but lowers the spatial resolution leading to less reliable bar property measurements. The most distant radial bin may contain less than $N_{\rm part}$, but in practice this bin is never considered in our analysis. Following \citet{Dehnen2023}, after determining the initial number of bins, $N_{\rm bin}$, we add an additional $N_{\rm bin} - 1$ intermediate bins between the medians of all initial bins, leaving us with $2N_{\rm bin}-1$ overlapping bins in total. Within each bin, we calculate the $A_{2}(R)$ and $\phi_{2}(R)$ used in this work. 

In Fig.~\ref{fig:example_bars}, we show the face-on surface mass density projection, $\Sigma_{\star}$, of a barred galaxy, as well as the radial $A_{2}(R)$ and $\phi_{2}(R)$ profiles in the top and bottom rows of the left panel. The bar length, $R_{\rm bar}$, is shown as a solid blue line/circle, and is identified as the region around $A_{2}^{\rm max}$ (dashed red line/circle) within which $\phi_{2}(R)$ deviates from $\phi_{2}^{\rm max}$ by $\leq \pm 10^{\circ}$ while $A_{2}(R) \geq A_{2}^{\rm max}/2$, following the procedure outlined in \citet{Dehnen2023}. This region therefore has both an inner and outer radius, labelled $R_{0}$ and $R_{1}$ and shown as dotted and solid vertical blue lines /circles in Fig.~\ref{fig:example_bars}; when quoting a value for the bar radius we simply quote the \textit{outer} radius, i.e., $R_{1} = R_{\rm bar}$. We require $R_{\rm bar} > 1\,{\rm kpc}$, otherwise we do not identify a bar regardless of the value of $A_{2}^{\rm max}$. The face-on projection of the example galaxy shows that this method produces an $R_{\rm bar}$ which precisely captures the bars extent. Other methods can find similar bar lengths \citep[e.g.][]{Ghosh2024}, but we caution against assuming $R_{\rm bar} = R(A_{2}^{\rm max})$, which, as seen in Fig.~\ref{fig:example_bars}, produces a severe underestimate. 

Measuring the bar pattern speed is slightly more complicated. First, note that $\phi_{2}$ accurately determines a bar's orientation (evidenced by Fig.~\ref{fig:example_bars}, where we have aligned the bar with the x-axis). Second, the bar pattern speed, $\Omega_{\rm bar}$, can be determined instantaneously from the time derivative of $\phi_{2}$,
\begin{equation}
    \Omega_{\rm bar} = \frac{d \phi_{2}}{d t} = \sum_{j} \frac{\partial \phi_{2}}{\partial x_{j}} \frac{d x_{j}}{dt},
\end{equation}
where $x_{j}$ are the particle positions. However, \citet{Dehnen2023} noted critically that the net particle flux on the radial bins must be considered when calculating this derivative, otherwise $\Omega_{\rm bar}$ will have systematic errors of $5 - 20$ per cent. This bias can be minimised by weighing stellar particles with a window function, such as
\begin{equation}
    W(R) = (1 - Q)^2(1 + 2Q), {\rm where \,} Q = \frac{R^2 - R_{\rm m}^2}{R_{\rm e}^2 - R_{\rm m}^2},
\end{equation}
and $R_{\rm med}$ is the middle of the bar region, while $R_{\rm e} = R_{0}$ if $R < R_{\rm med}$ and $R_{\rm e} = R_{1}$ otherwise. Applying this window function to the initial Fourier decomposition, we get
\begin{equation}
    \mathcal{A_{\rm m}} \equiv \frac{\sum_j W(x_{j}) M_{j} e^{mi\theta_{j}}}{\sum_j M_{j}},
\end{equation}
such that the bar strength is,
\begin{equation}
    |A_{\rm m}|= \frac{\sqrt{B_{m}^2 + C_{m}^2}}{\sum_{j} M_{j}}. 
\end{equation}
Here, 
\begin{align}
    B_{m} &= \sum_{j} M_{j} W(x_{j}) \sin(m \theta_{j}), \text{\,and,}\\
    C_{m} &= \sum_{j} M_{j} W(x_{j}) \cos(m \theta_{j}). 
\end{align} 
Therefore, $\phi_{2}$ and $\Omega_{\rm bar}$ can be expressed by setting $m=2$ as
\begin{equation}
    \phi_{2} = \arg(\mathcal{A}_{\rm 2}) = \frac{1}{2} \arctan \left( \frac{B_{2}}{C_{2}} \right)
\end{equation}
and
\begin{equation}
    \Omega_{\rm bar} = \frac{d\phi_{2}}{dt} = \frac{C_{2} \dot{C_{2}} - B_{2}\dot{B_{2}}}{2(C_{2}^2 + B_{2}^{2})}, 
\end{equation} 
where $\dot{B_{2}}$ and $\dot{C_{2}}$ are the time derivatives of $B_{2}$ and $C_{2}$, respectively. This provides a simple way to measure the instantaneous bar pattern speed for any galaxy at any snapshot. We point the reader to \citet{Dehnen2023} for more details. 

\section{Alternative methods of determining the bar formation pathway}\label{apx:altmethods}
\begin{figure*}
	\includegraphics[width=\textwidth]{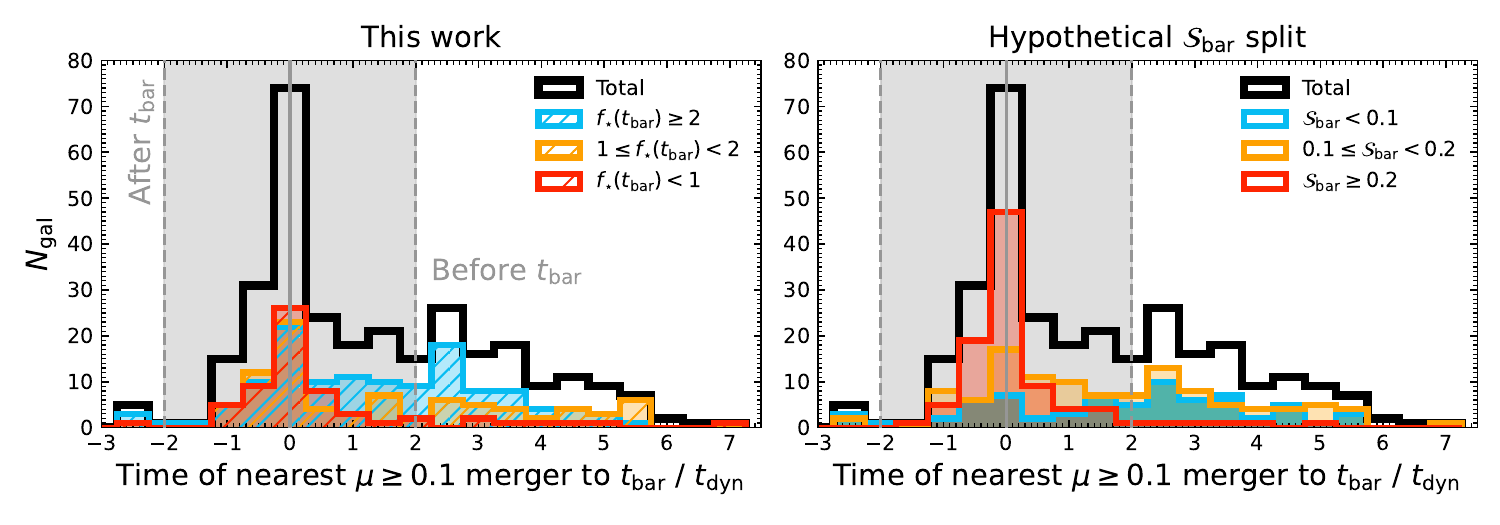}
    \caption{Histogram of the time relative to $t_{\rm bar}$, normalised by $t_{\rm dyn}$, of the nearest merger with stellar mass ratio $\mu \geq 0.1$. 
    The total distribution is show as a thick black line. Histograms for bars formed in different $f_{\star}(t_{\rm bar})$ and different $\mathcal{S}_{\rm bar}$ are shown in the left and right panels, respectively.}
    \label{fig:other_methods}
\end{figure*}
\begin{figure}
	\includegraphics[width=\columnwidth]{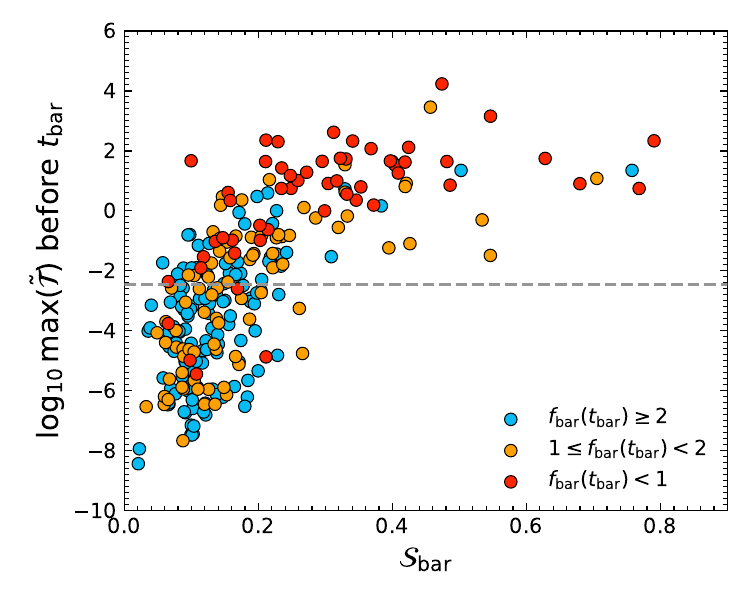}
    \caption{Scatter between $\max(\tilde{\mathcal{T}})$ and $\mathcal{S}_{\rm bar}$ for galaxies in our sample that form one bar episode. Points are coloured according to $f_{\star}(t_{\rm bar})$, as before. The horizontal grey dashed line denotes a threshold previously used to delineate tidal and secular bars \citep{Ansar2023, Lopez2024}. }
    \label{fig:other_methods2}
\end{figure}

In Fig.~\ref{fig:other_methods} and Fig.~\ref{fig:other_methods2} we show two alternative methods to separate bar formation episodes into ``secular'' and ``tidal'' bars -- definitions which we have shown to be largely arbitrary in a cosmological context. However, for the sake of comparison, we discuss below whether arbitrary thresholds in our tidal strength parameter, $\mathcal{S}_{\rm bar}$, can agree with the results of previous methods. 

In Fig.~\ref{fig:other_methods}, we compare our results with the method used in \citet{RosasGuevara2024}, which follows the galaxy merger trees to determine if a merger occurs near the time of bar formation. \citet{RosasGuevara2024} identify the time of the closest merger to $t_{\rm bar}$ along the main progenitor branch with a stellar mass ratio $\mu \geq 0.1$, where $\mu = M_{\star}^{(2)}/M_{\star}^{(1)}$, and $M_{\star}^{(1)}$ is the stellar mass of the more massive galaxy \citep[see][for details on characterising the merger histories of \textsc{TNG50} galaxies]{Bottrell2024}, and search for such mergers within $t_{\rm bar}\pm 2t_{\rm dyn}$. Galaxies which experience these mergers within this time window form ``tidal bars''. They also search for flybys within $t_{\rm bar}\pm 2t_{\rm dyn}$ with mass ratio $\mu \geq 0.1$ that pass within $100\,{\rm kpc}$ of the bar forming galaxies; these are also labelled as ``tidal bars''. All other the bar formation events would be labelled as ``secular bars''. 

In the upper right panel of Fig.~\ref{fig:other_methods}, we see how this definition compares against our selection on $f_{\star}(t_{\rm bar})$ used in this work, where we show histograms of the time of the nearest $\mu\geq0.1$ merger (normalised by $t_{\rm dyn}$) for all galaxies in our sample as a thick black line, and the distribution of our $f_{\star}(t_{\rm bins})$ as coloured, hatched histograms. The shaded region indicates the window within which a merger must occur for the bar to be tidal by the \citet{RosasGuevara2024} definition. A clear spike in mergers at $t_{\rm bar}$ indicates the importance of merger-driven tidal interactions in bar formation, but the long tail in the distribution also shows that many bars have $\mu \geq 0.1$ mergers far before $t_{\rm bar}$. There is an approximately equal number of bars in the different $f_{\star}(t_{\rm bar})$ bins that form in the same snapshot as a $\mu\geq0.1$ merger, as expected from Fig.~\ref{fig:Ta_MsMh_distribution}. However, there are far more highly stellar-dominated bars with nearest mergers far before $t_{\rm bar}$, and nearly no bar forming DM-dominated galaxies in the same range, indicating marginal agreement between our selections and the \citet{RosasGuevara2024} definitions; this should be no surprise, as these definitions are not related. 

A fairer comparison is shown in the upper right panel, where we show how three arbitrary $\mathcal{S}_{\rm bar}$ thresholds agree with the \citet{RosasGuevara2020} definitions: all galaxies with $\mathcal{S}_{\rm bar} \geq 0.2$ form their bars within $\pm2t_{\rm dyn}$ of a $\mu\geq0.1$ merger. In contrast, far fewer bars with $\mathcal{S}_{\rm bar} < 0.1$ and $0.1 \leq \mathcal{S}_{\rm bar} < 0.2$ form during this window. Ultimately, we choose not to use this method because there is no clear time window or $\mathcal{S}_{\rm bar}$ that can delineate tidal and secular bar formation, and there is no reason that $\mu\geq0.1$ mergers \textit{must} lead to bar formation, particularly if the tidal forces during the merger are particularly small once accounting for the rest of the nearby environment. Furthermore, the choice of $\mu \geq 0.1$ is highly arbitrary; \citet{Cavanagh2022} indicated that bar formation can be associated to mergers with $\mu$ far above and below this. 

An alternative method uses the scaled tidal index, the scalar $\tilde{\mathcal{T}}$ \citep[derived in Eq.~\ref{eq:simpleT}, but see e.g.][for a recent application]{Ansar2023}. $\tilde{\mathcal{T}}$ measures a simplified tidal force on the bar forming galaxy individually for every satellite in the same halo as the bar forming galaxy, for each snapshot within one halo $t_{\rm dyn}$ before $t_{\rm bar}$. When calculating the scalar $\tilde{\mathcal{T}}$, we choose $m = M_{\rm sat}$ and $r = D_{\rm sat}$, the total bound mass of any satellite and the distance between the centres of potential of our barred galaxies and the associated satellite, respectively. We again choose $r_{\rm t} = 4r_{\star, 1/2}$ as the outer radius enclosing the mass $M_{\rm t}(<r_{\rm t})$ of our barred galaxy. 

In Fig.~\ref{fig:other_methods2}, we show how $\max(\tilde{\mathcal{T}})$ measured before $t_{\rm bar}$ compares to our $\mathcal{S}_{\rm bar}$ metric. Galaxies are coloured by our $f_{\star}(t_{\rm bar})$ selections, respectively. \citet{Ansar2023} and \citet{Lopez2024} used $\max(\tilde{\mathcal{T}}) \gtrsim -2.45$ to divide the tidal and secular bars, in part because \citet{Purcell2011} showed that satellite interactions with this value of $\max(\tilde{\mathcal{T}})$ can induce a strong bar in a MW-mass galaxy; we show this as a horizontal dashed grey line. Bars in our sample form during a variety of $\max(\tilde{\mathcal{T}})$, but galaxies with $\mathcal{S}_{\rm bar} < 0.1$ are found at low values, and galaxies with $\mathcal{S}_{\rm bar} \geq 0.2$ are found at higher values, often above $\max(\tilde{\mathcal{T}}) = -2.45$. Some overlap exists in this parameter space, but primarily from galaxies with $0.1 \leq \mathcal{S}_{\rm bar} < 0.2$, as expected. Our reasons to not use $\max(\tilde{\mathcal{T}})$ are documented in Section~\ref{sec:tides}.  

\section{Bar properties at the time of bar formation} \label{apx:bar_props}
\begin{figure}
    \includegraphics[width=\columnwidth]{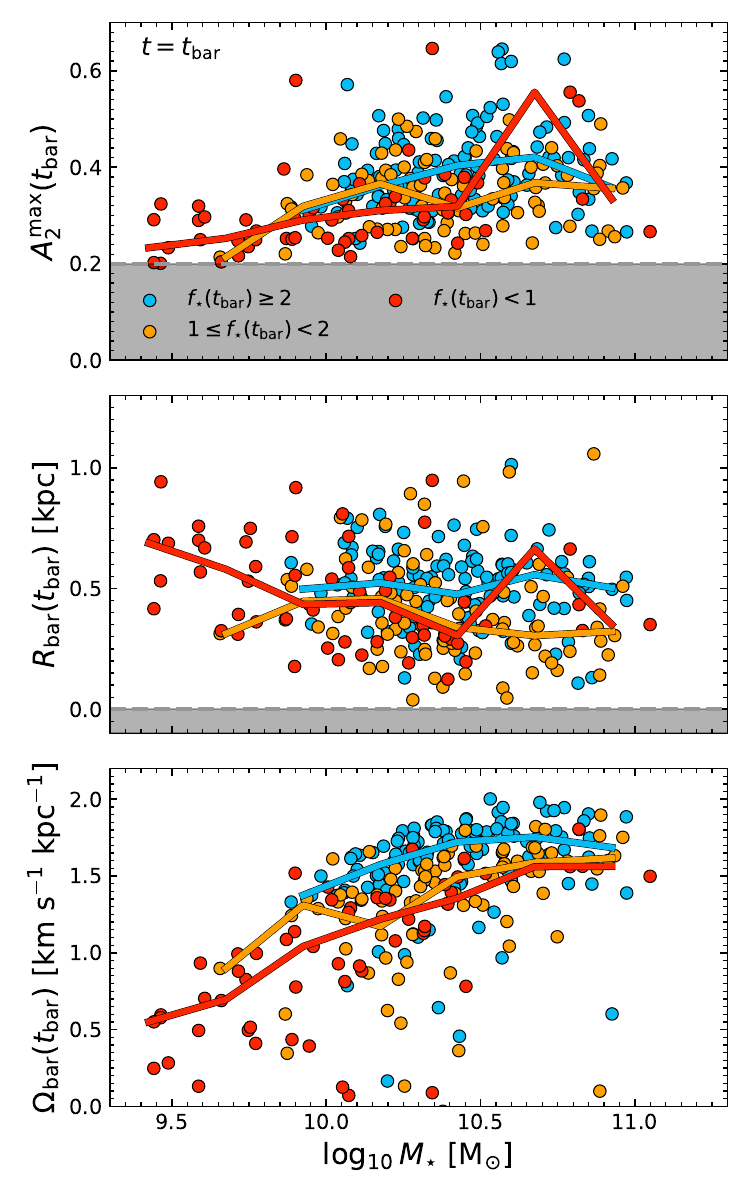}
    \caption{From top to bottom we show the bar strengths, lengths, and pattern speeds as a function of their stellar mass at the time of bar formation, $t_{\rm bar}$. Points are coloured according to whether $f_{\star}(t_{\rm bar})$, as before. Lines of the same colour indicate the median value of these populations as a function of stellar mass. Shades regions indicate the limits we require for a bar to have formed.}
    \label{fig:bar_props_z0}
\end{figure}
While we show the time evolution of normalised bar properties in Fig.~\ref{fig:bar_props_evo}, it can also be illuminating to view the un-normalised properties of bars at the time of bar formation. In Fig.~\ref{fig:bar_props_z0}, we show the $t=t_{\rm bar}$ bar strength, length, and pattern speed, all in physical units, as a function of $M_{\star}$ measured at the same time. Galaxies are binned based on $f_{\star}(t_{\rm bar})$, and coloured as before. Thick lines of the same colour indicate sub-sample medians as a function of $M_{\star}$. 

At the time of bar formation, bars that form in highly stellar dominated discs (often through secular evolution), tend to be stronger, longer, and more quickly rotating at fixed mass when compared to bars that form in less stellar-dominated discs. Galaxies that form bars in DM-dominated discs (i.e., tidal bars), are among the weakest, and most slowly rotating. However, these same galaxies are also particularly long, especially when found at stellar masses below $M_{\star} \leq 10^{10}\,{\rm M_{\odot}}$. 

\bsp	
\label{lastpage}
\end{document}